\pdfoutput=1
\documentclass[twocolumn,prb,superscriptaddress,groupaddress]{revtex4-1}
\usepackage[T1]{fontenc}
\usepackage[latin9]{inputenc}
\usepackage{color}
\usepackage{amsmath}
\usepackage{amssymb}
\usepackage{graphicx}

\makeatletter

\pdfpageheight\paperheight
\pdfpagewidth\paperwidth

\renewcommand\[{\begin{equation}}
\renewcommand\]{\end{equation}}
\usepackage{anysize}
\usepackage{graphicx}
\usepackage{dcolumn}% Align table columns on decimal point
\usepackage{bm}
\usepackage{slashed}
\usepackage{verbatim}
\usepackage{braket} %\Braket, \Bra, \Ket
\usepackage{longtable}
\usepackage{multirow}
\usepackage{amsmath}
\usepackage{xcolor}
\usepackage{amssymb}
\usepackage{appendix}
\usepackage{wrapfig}
\usepackage{epstopdf}
\usepackage{hyperref}              %Creates hyperlinks in cross references
\hypersetup{
    colorlinks=true,
    linkcolor=blue,
    filecolor=magenta,      
    urlcolor=magenta,
    citecolor=violet,
}

\newcommand{\up}{\uparrow}
\newcommand{\dw}{\downarrow}
\marginsize{1.5cm}{1.5cm}{1cm}{1cm}

\makeatother

\begin{document}

\title{Coherent long-range transfer of two-electron states in ac driven
triple quantum dots}

\author{Jordi Pic\'o-Cort\'es}
\email{jordi.p.cortes@csic.es}
\affiliation{Materials Science Factory, Instituto de Ciencia de Materiales de
Madrid (CSIC), 28049, Madrid, Spain }
\affiliation{Institute for Theoretical Physics, University of Regensburg, 93040 Regensburg, Germany}

\author{Fernando Gallego-Marcos}

\author{Gloria Platero}
\affiliation{Materials Science Factory, Instituto de Ciencia de Materiales de
	Madrid (CSIC), 28049, Madrid, Spain }

\begin{abstract}
Preparation and transfer of quantum states is a fundamental task in
quantum information. We propose a protocol to prepare a state in the
left and center quantum dots of a triple dot array and transfer it
directly to the center and right dots. Initially the state in the
left and center dots is prepared combining the exchange interaction
and magnetic field gradients. Once in the desired state, ac gate voltages
in the outer dots are switched on, allowing to select a given photoassisted
long-range path and to transfer the prepared state directly from one
edge to the other with high fidelity. We investigate the effect of
charge noise on the protocol and propose a configuration in which
the transfer can be performed with high fidelity. Our proposal can
be experimentally implemented and is a promising avenue for transferring
quantum states between two spatially separated two-level systems.
\end{abstract}
\maketitle

\section{Introduction}

Since the proposal by Cirac and Zoller to use photons for quantum
state transfer between atoms located at spatially-separated nodes
of a quantum network\cite{Cirac1997}, different works have explored
how to transfer a quantum state in optical~\cite{Vermersch2017}
and solid state devices\cite{McNeil2011,He2017}. Quantum dot arrays
have shown to be ideal solid state systems for hosting charge and
spin qubits\cite{Hanson2007}. Manipulation of qubits in GaAs semiconductor
double quantum dots has been exhaustively investigated\cite{Bluhm2011,Granger2012,Sanchez2013}.
Recently, experimental implementation of quantum dot arrays with increasing
number of dots has allowed to study new phenomena\cite{Ito2016,Zajac2016,Fujita2017},
such as geometrical frustration in triple quantum dots \cite{Korkusinski2007},
dynamical channel blockade\cite{Kotzian2016}, or the coherent control\cite{Gaudreau2011}
and state tomography\cite{Medford2013} of three spin states in triple
quantum dots\cite{Russ2017}.The implementation of direct quantum
state transfer between distant sites in quantum dot arrays is of great
interest for quantum information purposes. Long-range charge and spin
transfer, where the transfer occurs between non directly coupled distant
sites, has been demonstrated in arrays of three quantum dots\cite{Busl2013,Braakman2013,Sanchez2014,Wang2017}
and several proposals exist to extend long-range coupling to longer
arrays\cite{Ban2018,Rahman2010}. Recently it has been shown that
applying ac gate voltages new features in the current occur, such
as long range photoassisted charge\cite{Schreiber2011,GallegoMarcos2015},
energy and heat currents\cite{GallegoMarcos2017}, or current blockade
due to destructive interferences between virtual and real photoassisted
quantum paths\cite{GallegoMarcos2016}.

\begin{figure}
\begin{centering}
\includegraphics[width=1\columnwidth]{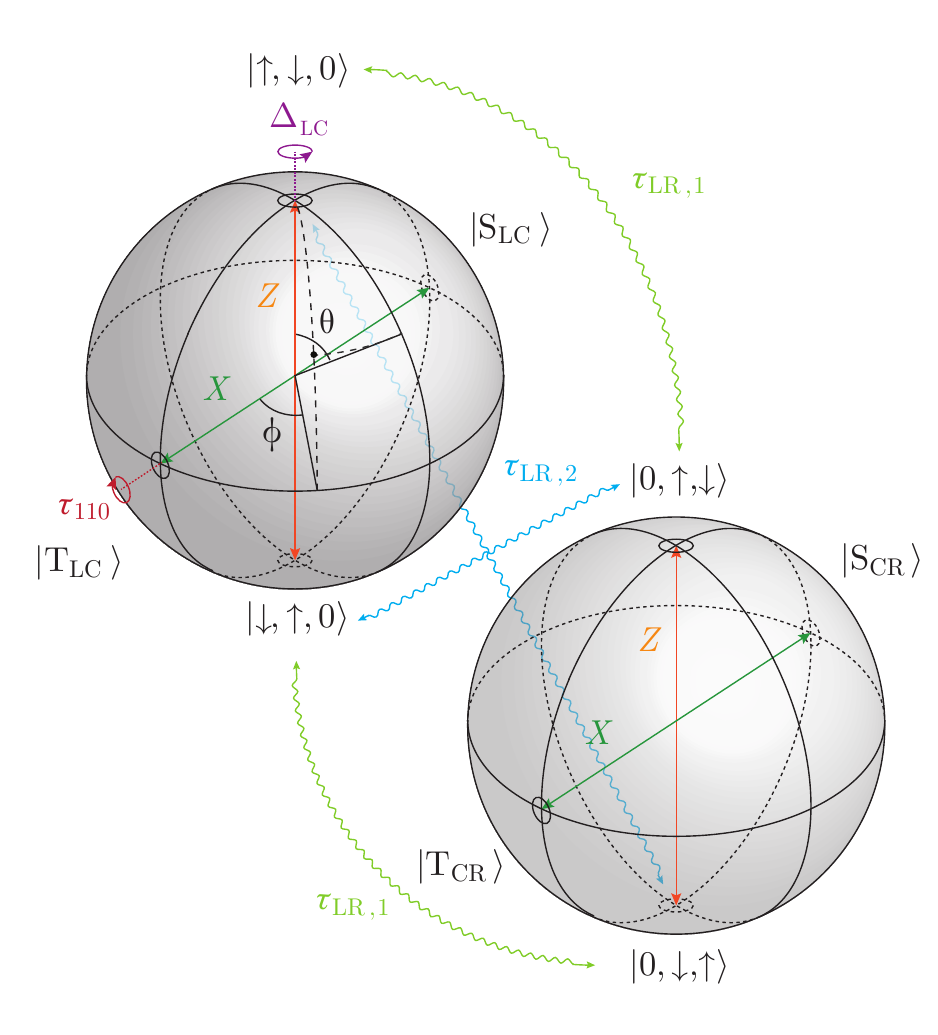}
\par\end{centering}
\caption{Scheme of the quantum state manipulation and transfer. The state is
prepared in the left two-level system defined in the left and center
quantum dots. Here it is represented as a Bloch sphere with angles
$\theta,\phi$. The angle $\theta$ is set through the exchange interaction
$\tau_{110}$, while $\phi$ is set through a magnetic field gradient
between the left and center dots, $\Delta_{\mathrm{LC}}$. The prepared
state is then transferred to the two-level system defined in the center
and right quantum dots through the long-range photoassisted paths
$\tau_{\mathrm{LR,1}}$~and~�$\tau_{\mathrm{LR,2}}$, denoted by
curly arrows.\label{fig:esquema}}
\end{figure}

Two-electron states in two quantum dots offer a flexible and well-studied
platform for quantum information purposes, forming the basis of the
well-known singlet-triplet qubit\cite{Hanson2007}. Combining electric
and magnetic control through the exchange interaction and magnetic
field gradients provides full single-qubit manipulation capabilities
and can be extended to include two-qubit operations\cite{Wardrop2014}.
The possibility of state transfer between spin-triplet qubits offers
new possibilities for the development of new quantum architectures
based on this platform. In that direction, a long-range protocol based
on a singlet-triplet qubit has been proposed recently\cite{Feng2018}
based on adiabatic transfer and Coulomb interaction engineering. 

In this work we propose how to prepare a quantum state with two electrons
in the left and center quantum dots of a triple quantum dot (TQD)
system and how to transfer it coherently to the center and right dots
by using ac gate voltages. The ac driving allows us to stop the evolution
of the prepared state and to select a long range quantum transfer
path. The two electrons are transferred simultaneously and coherently
with high fidelity, even in the presence of charge noise. Furthermore,
we develop a general transfer protocol for arbitrary gradient configurations,
ensuring that our proposal can be extended to longer quantum dot arrays.
The paper is organized as follows. In Section~\ref{sec:Theoretical-Model}
we introduce the effective Hamiltonian that we employ to study the
ac response of the system. In Section~\ref{subsec:woGrads} we propose
a transfer protocol in the case in which there are no magnetic field
gradients. In Section~\ref{subsec:withGrads}, we analyze the role
of magnetic gradients in the transfer process. Finally, in Section~\ref{subsec:Relaxation-and-decoherence}
we analyze the fidelity of the protocol under the effect of charge
noise and discuss other possible sources of decoherence. 

\section{Results}

\subsection{Theoretical model\label{sec:Theoretical-Model}}

We consider up to two electrons in a TQD in series. A external magnetic
field produces a Zeeman splitting within each dot. Two oscillating
electric field voltages are locally applied to the left and right
quantum dots $H_{\text{ac}}(t)=V_{\text{ac}}^{\text{L}}\cos(\omega t)\hat{\text{n}}_{\text{L}}+V_{\text{ac}}^{\text{R}}\cos(\omega t)\hat{\text{n}}_{\text{R}}$.
The Hamiltonian can be written in the interaction picture as $H_{I}(t)=\mathcal{U}_{I}(t)[H(t)-i\hbar\partial_{t}]\mathcal{U}_{I}^{\dagger}(t)$
where $\mathcal{U}_{I}(t)=\exp\left[(i/\hbar)\int H_{\mathrm{ac}}(t)dt\right]$.
Then, the Hamiltonian reads
\begin{align}
H_{I}(t) & =\sum_{i,\sigma}\epsilon_{i}\hat{\text{c}}_{i,\sigma}^{\dag}\hat{\text{c}}_{i,\sigma}+\sum_{i}B_{z,i}\hat{S}_{z,i}\nonumber \\
 & +\sum_{i<j,\sigma,\sigma'}U_{ij}\hat{n}_{i,\sigma}^{\dag}\hat{n}_{j,\sigma'}+\sum_{i}U_{ii}\hat{n}_{i,\up}^{\dag}\hat{n}_{i,\dw}\nonumber \\
 & +\sum_{\sigma}\sum_{\nu=-\infty}^{\infty}t_{\text{LC}}^{\nu}(t)(\hat{c}_{\text{L},\sigma}^{\dag}\hat{c}_{\text{C},\sigma}+\text{H.c.}),\nonumber \\
 & +\sum_{\sigma}\sum_{\nu=-\infty}^{\infty}t_{\text{CR}}^{\nu}(t)(\hat{c}_{\text{R},\sigma}^{\dag}\hat{c}_{\text{C},\sigma}+\text{H.c.})\label{eq:Hrot}
\end{align}
where $i,j=\{\text{L,C,R}\}$ and $\sigma,\sigma'=\{\up,\dw\}$. The
different parameters correspond to the on-site energy $\epsilon_{i}$,
and the Zeeman splitting $B_{z,i}$ of the $i$th dot; the inter-dot
interaction $U_{ij}$, the intra-dot interaction $U_{ii}$, and the
renormalized tunnel couplings between the dots $t_{\text{LC}}^{\nu}(t)=\tau_{\text{LC}}J_{\nu}(V_{\text{ac}}^{\text{L}}/\hbar\omega)\text{e}^{i\nu\omega t}$
and $t_{\text{CR}}^{\nu}(t)=\tau_{\text{CR}}J_{\nu}(V_{\text{ac}}^{\text{R}}/\hbar\omega)\text{e}^{i\nu\omega t}$,
where $J_{\nu}(\alpha)$ is the $\nu$th Bessel function of the first
kind. We also denote the energy of each state as $E_{ij}=\epsilon_{i}+\epsilon_{j}+U_{ij}$.
We assume a configuration where the energy differences of $\ket{\sigma,0,\sigma'}$
and the doubly-occupied states with the states $\ket{\sigma,\sigma',0}$
and $\ket{0,\sigma,\sigma'}$ are the largest energy scales in the
system, i.e., $\{V_{\text{ac}}^{\text{L}},V_{\text{ac}}^{\text{R}},\hbar\omega,|\tau_{ij}|,|E_{\text{LC}}-E_{\text{CR}}|,|\Delta_{ij}|\}\ll\{|\delta_{101}|,|\delta_{020}|,|\delta_{200}|,|\zeta_{101}|,|\zeta_{020}|,|\zeta_{002}|\}$,
where $\delta_{101}\equiv E_{\text{LR}}-E_{\text{LC}}$, $\delta_{020}\equiv E_{\text{CC}}-E_{\text{LC}}$,
$\delta_{200}\equiv E_{\mathrm{LL}}-E_{\mathrm{LC}}$, $\zeta_{101}\equiv E_{\text{LR}}-E_{\text{CR}}$,
$\zeta_{020}\equiv E_{\text{CC}}-E_{\text{CR}}$, $\zeta_{002}\equiv E_{\mathrm{RR}}-E_{\mathrm{CR}}$
and~$\Delta_{ij}=(B_{z,j}-B_{z,i})/2$. In this regime we obtain
an effective Hamiltonian with virtual tunneling as the leading order
of perturbation by means of a time-dependent Schrieffer-Wolff transformation\cite{Goldin2000}.
Written in the basis $\{\ket{\up,\dw,0},\ket{\dw,\up,0},\ket{0,\up,\dw},\ket{0,\dw,\up}\}$,
the effective Hamiltonian reads
\begin{equation}
H_{\mathrm{eff}}(t)=\left[\begin{array}{cccc}
\tilde{E}_{|\uparrow,\downarrow,0\mathcal{i}}(t) & \tau_{110}^{*}(t) & \tau_{\mathrm{LR,1}}^{*}(t) & \tau_{\mathrm{LR,2}}^{*}(t)\\
\tau_{110}(t) & \tilde{E}_{|\downarrow,\uparrow,0\mathcal{i}}(t) & \tau_{\mathrm{LR,2}}^{*}(t) & \tau_{\mathrm{LR,1}}^{*}(t)\\
\tau_{\mathrm{LR,1}}(t) & \tau_{\mathrm{LR,2}}(t) & \tilde{E}_{|0,\uparrow,\downarrow\mathcal{i}}(t) & \tau_{011}^{*}(t)\\
\tau_{\mathrm{LR,2}}(t) & \tau_{\mathrm{LR,1}}(t) & \tau_{011}(t) & \tilde{E}_{|0,\downarrow,\uparrow\mathcal{i}}(t)
\end{array}\right].\label{eq:EffH}
\end{equation}

$\tilde{E}_{k}(t)$ are the renormalized energies of the states, $\tau_{110}(t)$
and $\tau_{011}(t)$ are the rates for the exchange interactions due
to virtual transitions through the doubly occupied states $\ket{\uparrow\downarrow,0,0},\ket{0,\uparrow\downarrow,0}$
and $\ket{0,0,\uparrow\downarrow}$. $\tau_{\mathrm{LR,1}}(t)$ and
$\tau_{\mathrm{LR,2}}(t)$ are the amplitudes for the long-range processes
connecting $\{\ket{\up,\dw,0},\ket{\dw,\up,0}\}$ and $\{\ket{0,\up,\dw},\ket{0,\dw,\up}\}$
by virtual transitions through the $\ket{0,\uparrow\downarrow,0}$
and $\ket{\sigma,0,\sigma'}$ states. The expressions for the different
terms in the effective Hamiltonian are given in the Supplementary
information. 

The proposed protocol consists of the preparation of a state
\begin{equation}
|\Psi_{\mathrm{L}}\mathcal{i}=\cos(\theta_{\mathrm{L}}/2)\ket{\uparrow,\downarrow,0}+e^{i\phi_{\mathrm{L}}}\sin(\theta_{\mathrm{L}}/2)\ket{\downarrow,\uparrow,0}\label{eq:psiL}
\end{equation}
 in the two-level system $\mathcal{Q}_{\mathrm{L}}=\{\ket{\up,\dw,0},\ket{\dw,\up,0}\}$
defined in the left and center dots, where $\phi_{\mathrm{L}}$ can
be defined in terms of the density matrix $\rho$ as

\[
\phi_{\text{L}}=\text{Arg}\left[\frac{\braket{\up,\dw,0|\rho|\dw,\up,0}}{\sqrt{\braket{\up,\dw,0|\rho|\up,\dw,0}\braket{\dw,\up,0|\rho|\dw,\up,0}}}\right],
\]

Manipulation of both $\theta_{\mathrm{L}}$ and $\phi_{\mathrm{L}}$
is attained by a combination of the magnetic field gradients and the
exchange interaction due to virtual processes involving the doubly
occupied states, with corresponding transition rates $\tau_{110}(t)$
and $\tau_{011}(t)$. Then, the prepared state can be transferred
to the two-level system $\mathcal{Q}_{\mathrm{R}}=\{\ket{0,\up,\dw},\ket{0,\dw,\up}\}$
defined in the center and right dots, yielding 

\[
\ket{\psi_{\mathrm{R}}}=\cos(\theta_{\mathrm{L}}/2)\ket{0,\up,\dw}+e^{i\phi_{\mathrm{L}}}\sin(\theta_{\mathrm{L}}/2)\ket{0,\dw,\up}.
\]

This transfer is carried out through the long-range photoassisted
paths, with rates given by $\tau_{\mathrm{LR,1}}(t)$~and~$\tau_{\mathrm{LR,2}}(t)$.
The former, $\tau_{\mathrm{LR,1}}(t)$, connects states in the same
poles of the Bloch sphere, while the latter, $\tau_{\mathrm{LR,2}}(t)$
connects states in opposite poles of the sphere (see Fig.~\ref{fig:esquema}).
Two problems arise from this configuration. First, the exchange interactions
act on the quantum state during the transfer. Second, there are two
different transference channels, which limits the fidelity. Both can
be solved by using ac-driving fields. By choosing the proper ac-driving
amplitudes, the interference between the different photoassisted paths
with rates $\tau_{110}(t),\tau_{011}(t),$ and~$\tau_{\mathrm{LR,2}}(t)$
can be used to nullify these processes, as will be shown below.

We consider in Sec.~\ref{subsec:woGrads} the simpler case in which
there are no magnetic field gradients. Then, we will consider the
general case with arbitrary magnetic field gradients in Sec.~\ref{subsec:withGrads}.

\subsection{Without magnetic field gradients\label{subsec:woGrads}}

\begin{figure}
\begin{centering}
\includegraphics[width=1\columnwidth]{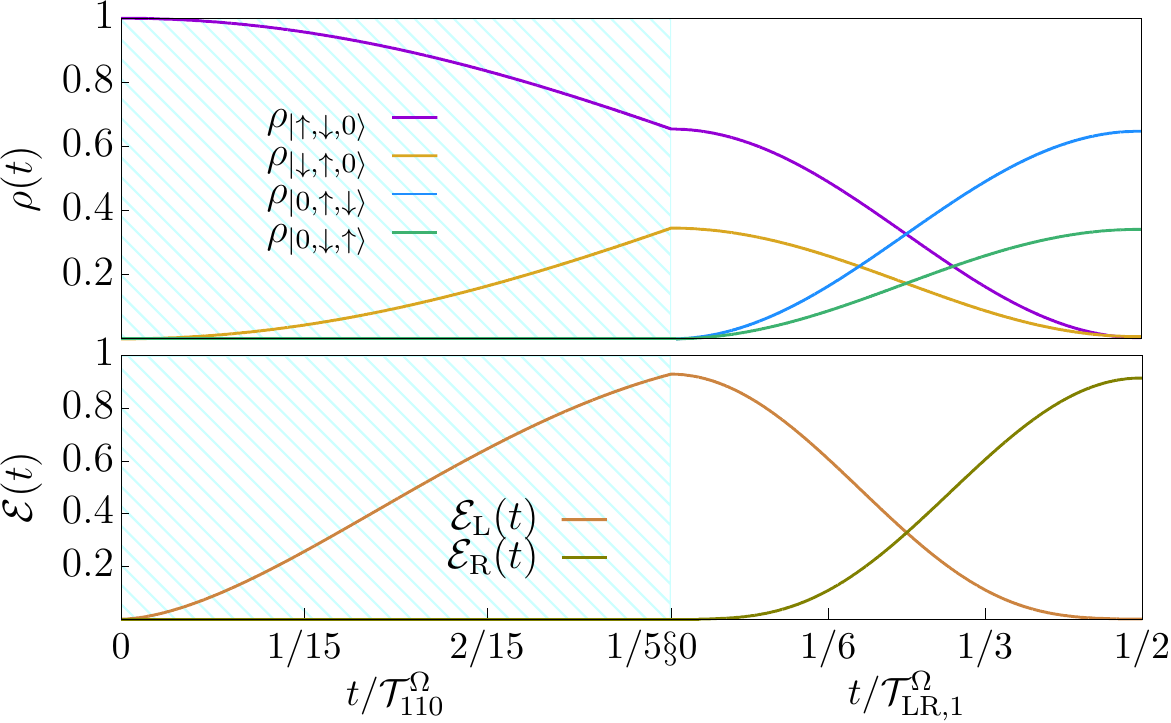}
\par\end{centering}
\caption{(top) Time evolution of the population of the different states. (bottom)
Time evolution of the entanglement between the spins. The two spins
in $\mathcal{Q}_{\mathrm{L}}$ are prepared into an state $|\Psi_{\mathrm{L}}\mathcal{i}=\cos(\theta_{\mathrm{L}}/2)\ket{\uparrow,\downarrow,0}+e^{i\phi_{\mathrm{L}}}\sin(\theta_{\mathrm{L}}/2)\ket{\downarrow,\uparrow,0}$
with $\ensuremath{\theta_{\text{L}}=2\pi/5}$ and $\phi_{\mathrm{L}}=\pi/2$
by means of $\tau_{110}$ (dashed green area) and then the state is
transferred to $\mathcal{Q}_{\mathrm{R}}$ (white area). Parameters:
$\tau_{\text{LC}}=\tau_{\text{CR}}=\tau=30\mu\text{eV}$, $\epsilon_{\text{C}}/\omega=6.48$,
$\tilde{E}_{\text{LC}}=0$ and $\tilde{E}_{\text{CR}}=\omega$. In
the left part of the figure (blue dashed area) the ac gate voltages
are switched off. In the right part of the figure: $V_{\text{ac}}^{\text{L}}=3.94\mu\text{eV}$
and $V_{\text{ac}}^{\text{R}}=3.57\mu\text{eV}$, $\omega=10\tau$.\label{fig:entangTransfer}}
\end{figure}

Our first protocol consists on preparing a quantum state in $\mathcal{Q}_{\mathrm{L}}$
allowing only $\theta_{\mathrm{L}}$ to evolve (see Eq.~\ref{eq:psiL})
and then transferring it to $\mathcal{Q}_{\mathrm{R}}$. The procedure
can be fashioned as an entanglement generation between the single
spins in $\mathcal{Q}_{\mathrm{L}}$ dots and a transfer of the entangled
spins to $\mathcal{Q}_{\mathrm{R}}$. Initially, we turn the ac voltages
off and assume that there is no charge transfer between $\mathcal{Q}_{\mathrm{L}}$
and $\mathcal{Q}_{\mathrm{R}}$. Under the assumptions leading to
Eq.~\ref{eq:EffH}, this requires that the energy difference between
the states in $\mathcal{Q}_{\mathrm{L}}$ and $\mathcal{Q}_{\mathrm{R}}$
is much larger than the amplitudes of the long-range rates $\tau_{\mathrm{LR,1}}$
and $\tau_{\mathrm{LR,2}}$.

The initial state is taken as $\ket{\up,\dw,0}$. The desired value
of $\theta_{\mathrm{L}}$ can be set by allowing the system to evolve
by means of the virtual transitions with the doubly occupied states
$\ket{\up\dw,0,0}$ and $\ket{0,\up\dw,0}$ through $\tau_{110}$,
yielding the state $|\Psi_{\mathrm{L}}\mathcal{i}=\cos(\theta_{\mathrm{L}}/2)\ket{\uparrow,\downarrow,0}+e^{i\pi/2}\sin(\theta_{\mathrm{L}}/2)\ket{\downarrow,\uparrow,0}$.
With the ac voltages turned off, $\tau_{110}$ is given by
\begin{align}
\tau_{110} & =-\frac{\tau_{\mathrm{LC}}^{2}}{2}\left(\frac{1}{\delta_{200}}+\frac{1}{\delta_{020}}\right)\label{eq:tau110-no-ac}
\end{align}

This process has a Rabi period $\mathcal{T}_{110}^{\Omega}=\pi\hbar/|\tau_{110}|$
which can be controlled either by modifying the detuning between the
left and center dots (i.e: controlling $\delta_{020}$) or by symmetric
control of the tunneling barriers\cite{Russ2017} (i.e: controlling
$\tau_{\mathrm{LC}}$). The latter method has the benefit of allowing
for operation under the sweetspot condition\cite{Fei2015,Martins2016},
resulting in lower sensitivity to charge noise. 

Once the spins are in the desired state, the ac voltages are turned
on and the state is transferred to the center and right dots. With
the ac voltages on, the diagonal terms of Eq.~\ref{eq:EffH} are
time-dependent. To obtain the resonance condition that allows us to
transfer the state, we calculate the mean in time of the diagonal
terms, the \emph{mean energies.} These can be obtained as\footnote{See Supplementary information.}

\begin{align}
\tilde{E}_{\mathrm{LC}} & =E_{\mathrm{LC}}\nonumber \\
&-\sum_{\nu}\left[\frac{|t_{\mathrm{CR}}^{\nu}|^{2}}{\delta_{101}+\nu\hbar\omega}+\frac{|t_{\mathrm{LC}}^{\nu}|^{2}}{\delta_{020}-\nu\hbar\omega}+\frac{|t_{\mathrm{LC}}^{\nu}|^{2}}{\delta_{200}-\nu\hbar\omega}\right]\label{eq:ELCren}
\end{align}

\begin{align}
\tilde{E}_{\mathrm{CR}} & =E_{\mathrm{CR}}\nonumber\\ &-\sum_{\nu}\left[\frac{|t_{\mathrm{LC}}^{\nu}|^{2}}{\zeta_{101}+\nu\hbar\omega}+\frac{|t_{\mathrm{CR}}^{\nu}|^{2}}{\zeta_{020}-\nu\hbar\omega}+\frac{|t_{\mathrm{CR}}^{\nu}|^{2}}{\zeta_{002}-\nu\hbar\omega}\right]\label{eq:ECRren}
\end{align}

Here $\nu$ is the sideband index and goes from $-\infty$ to $\infty$
unless explicitly noted. Then, we assume that the difference between
the mean energies of the initial (left) and final (right) states is
$n\hbar\omega$. If $n=0$, the tunnel barrier between the center
and right dots has to be raised so that $\tau_{\mathrm{CR}}\simeq0$
while preparing the state avoiding electron transfer to the rightmost
dot. During the transfer, the tunnel barriers are then lowered to
allow the electrons to tunnel to the center and right dots. If $n\neq0$
and $\omega\gg\tau_{\mathrm{LC}},\tau_{\mathrm{CR}}$, $\mathcal{Q}_{\mathrm{L}}$
and $\mathcal{Q}_{\mathrm{R}}$ will only be coupled when the ac field
is turned on. This eliminates the need to manipulate the tunnel amplitudes
for the state transfer.

When the resonance condition $|\tilde{E}_{\mathrm{CR}}-\tilde{E}_{\mathrm{LC}}|=n\hbar\omega$,
$n\mathcal{2}\mathbb{N}$ is met, we can use the rotating wave approximation
(RWA), in which the energies of the Hamiltonian are shifted to the
desired resonance and the fast oscillating terms are neglected. For
that we apply a unitary transformation: $\mathcal{U}_{\mathrm{RWA}}^{\dag}(t)[H_{\text{eff}}(t)-i\hbar\partial_{t}]\mathcal{U}_{\mathrm{RWA}}(t)$
where $\mathcal{U}_{\mathrm{RWA}}(t)=\exp\left[-in\omega t(\hat{n}_{\ket{0,\up,\dw}}+\hat{n}_{\ket{0,\dw,\up}})\right]$.
This allows to obtain time-independent rates for the second order
processes, as given in the Supplementary information. Unless explicitly
noted, the formulas in the next sections are obtained from the RWA
approximation.

During the transfer process, the energy levels of $\ket{\up,\dw,0}$
and $\ket{\dw,\up,0}$ are resonant and virtual transitions between
the two states through the double occupied states modify $\theta_{\text{L}}$.
The formation of a dark state is required in order to stop the evolution
of $\theta_{\text{L}}$. Only if the state is a singlet or a triplet,
the state is an eigenstate of the exchange Hamiltonian, $\theta_{\mathrm{L}}$
does not change during the transfer process and ac fields are not
required to stop the evolution of $\theta_{\mathrm{L}}$. For a general
state, destructive interferences between the virtual photoassisted
paths may lead to $\tau_{110}^{\text{RWA}}=0$. This occurs for values
of the driving amplitude $V_{\text{ds}}^{\text{L}}$ such that
\begin{equation}
\sum_{\nu}J_{\nu}^{2}\left(\frac{V_{\text{ds}}^{\text{L}}}{\hbar\omega}\right)\left(\frac{1}{\delta_{020}-\nu\hbar\omega}+\frac{1}{\delta_{200}-\nu\hbar\omega}\right)=0\label{eq:darkState}
\end{equation}

Hence, the time evolution of $\theta_{\text{L}}$ can be stopped at
any desired point through the ac gates by setting $V_{\text{ac}}^{\text{L}}=V_{\text{ds}}^{\text{L}}$.
Similarly, for $\mathcal{Q_{\mathrm{R}}}$, a similar dark state condition
can be obtained for an ac driving amplitude $V_{\text{ac}}^{\text{R}}=V_{\text{ds}}^{\text{R}}$. 

There are two possible transport channels between $\mathcal{Q}_{\mathrm{L}}$
and $\mathcal{Q}_{\mathrm{R}}$ (see Fig.~\ref{fig:esquema}), controlled
by the virtual tunneling couplings $\tau_{\mathrm{LR,1}}^{n}$ and
$\tau_{\mathrm{LR,2}}^{n}$\cite{Note1}. Only if the state is a triplet,
transitions through the singlet $\ket{0,\up\dw,0}$ are forbidden
and $\tau_{\mathrm{LR,2}}^{n}=0$ always. For a general state, the
simultaneous presence of the two channels limits the fidelity of the
transfer process and the transition rate corresponding to one of the
long range photo-assisted paths, either $\tau_{\text{LR},1}^{n}$
or $\tau_{\text{LR},2}^{n}$, has to be set to zero. The ac voltage
can induce a destructive interference between the sidebands and nullify
$\tau_{\mathrm{LR,1}}^{n}$ or $\tau_{\mathrm{LR,2}}^{n}$ in the
same way as for $\tau_{110}$ and $\tau_{011}$. For concreteness,
we consider transfer between $\mathcal{Q}_{\mathrm{L}}$ and $\mathcal{Q_{\mathrm{R}}}$
 just through $\tau_{\text{LR},1}^{n}$. Then, $\tau_{\text{LR},2}^{n}$
is suppressed for a set of values 
\begin{equation}
\epsilon_{\text{C}}^{\text{ds}}=\left\{ \epsilon_{\text{C}}\ \big|\ \tau_{\text{LR},2}^{n}=0\ \&\ \tau_{\text{LR},1}^{n}\neq0\right\} ,\label{eq:eCds}
\end{equation}
where $\epsilon_{\text{C}}^{\text{ds}}$ is the energy of the central
level at which the destructive interference between the virtual photon-sidebands
occurs and $\tau_{\text{LR},2}^{n}=0$. In Fig.~\ref{fig:entangTransfer}
we have plotted the occupation of the relevant states and the entanglement
of the two spins during the preparation and transfer protocol. In
the blue dashed area, $\theta_{\mathrm{L}}$ is fixed by letting the
state evolve under $\tau_{110}$ for a certain time, with the ac voltages
turned off. Then, the ac voltages are turned on, connecting $\mathcal{Q}_{\mathrm{L}}$
to $\mathcal{Q}_{\mathrm{R}}$. In the white area, the state is transferred
through the $\tau_{\mathrm{LR,1}}^{n}$ process from $\mathcal{Q}_{\mathrm{L}}$
to $\mathcal{Q}_{\mathrm{R}}$. 

\subsection{With magnetic field gradients\label{subsec:withGrads}}

\begin{figure}
\begin{centering}
\includegraphics[width=1\columnwidth]{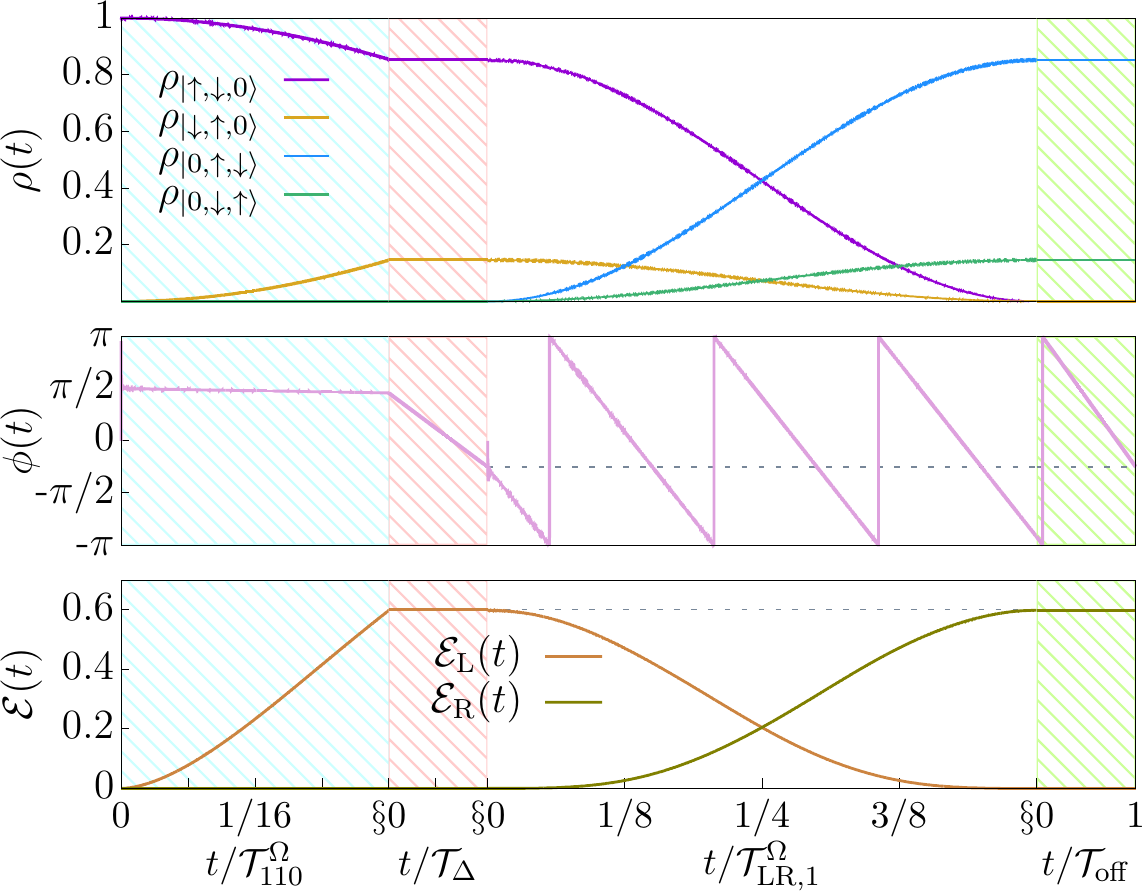}
\par\end{centering}
\caption{For $\Delta_{\mathrm{LC}}=\Delta_{\mathrm{CR}}$. (top) Time evolution
of the population of the states. (center) Time evolution of $\phi(t)$.
The phase $\phi(t)$ is defined in $\mathcal{Q}_{\mathrm{L}}$ during
the manipulation process and in \emph{$\mathcal{Q}_{\mathrm{R}}$
}during the transfer process. (bottom) Time evolution of the entanglement
$\mathcal{E}(t)$ between the two spins. From left to right: in the
blue dashed area $\theta_{\mathrm{L}}$ is fixed to the desired value
of $\theta_{\mathrm{L}}=\pi/4$; in the red dashed area the polar
angle $\phi_{\mathrm{L}}=-\pi/4$ is set through the magnetic field
gradient while $\tau_{110}\simeq0$; in the white area the two tunnel
barriers are lowered and the state is transferred through $\tau_{\mathrm{LR,1}}$;
in the green dashed area the phase is corrected to its value of $\phi_{\mathrm{L}}=-\pi/4$.
Parameters: $\Delta_{\mathrm{LC}}=\Delta_{\mathrm{CR}}=0.13$~$\mathrm{\mu eV}$.
$\tau_{\mathrm{LC}}=\tau_{\mathrm{CR}}=30$~$\mathrm{\mu eV}$ in
the white areas and the blue dashed areas and $\tau_{\mathrm{LC}}=\tau_{\mathrm{CR}}=0$
in the red and green dashed areas. $\delta_{020}=\zeta_{020}=4.25$~$\text{meV}$,
$\delta_{101}=\zeta_{101}=2.28$~$\text{meV}$, $n=0$. In the dashed
areas the ac gate voltages are switched off. In the right part of
the figure (white areas): $V_{\text{ac}}^{\text{L}}=V_{\mathrm{ac}}^{\mathrm{L}}=5.25\,\text{meV}$
$\omega=0.5$~$\mathrm{meV}$. The gray dashed lines are a visual
guide indicating the desired value of $\phi$ and the entanglement
of the initially prepared state. \label{fig:osRabi_LIN}}
\end{figure}

A magnetic field gradient, produced for instance by nanomagnets\cite{Petersen2013,Forster2015,Yoneda2015},
allows for the generation of any state in $\mathcal{Q}_{\mathrm{L}}$.
As long as $|\delta_{101}|,|\delta_{020}|,|\delta_{200}|\gg\tau_{\mathrm{LC}}$,
leakage into the $\ket{\sigma,0,\sigma'},\ket{\uparrow\downarrow,0,0}$
and $\ket{0,\up\dw,0}$ states can be kept minimal. At this point,
the TQD operates as a two-level system $\{\ket{\up,\dw,0},\ket{\dw,\up,0}\}$
with Hamiltonian
\begin{equation}
H_{\mathrm{LC}}=-\Delta\tilde{E}_{\mathrm{LC}}\hat{\sigma}_{\mathrm{LC}}^{z}+\tau_{110}\hat{\sigma}_{\mathrm{LC}}^{x},\label{eq:HLC}
\end{equation}
with $\hat{\sigma}_{\mathrm{LC}}^{i}$ the $i$th Pauli matrix in
$\mathcal{Q}_{\mathrm{L}}$, $i=x,z$ and
\begin{equation}
\Delta\tilde{E}_{\mathrm{LC}}=\frac{1}{2}\left(\tilde{E}_{|\downarrow,\uparrow,0\mathcal{i}}-\tilde{E}_{|\uparrow,\downarrow,0\mathcal{i}}\right).\label{eq:renorm-energy-split}
\end{equation}

 The ground state for $\tau_{110}\simeq0$ is given by $\ket{\up,\dw,0}$
due to the magnetic field gradient. This state, located in the north
pole of the Bloch sphere depicted in Fig.~\ref{fig:esquema} only
acquires a global phase as a result of the gradient, therefore providing
a suitable platform for initialization. The desired  state is then
prepared starting from $\ket{\up,\dw,0}$ by a combination of the
magnetic field gradient and the exchange interaction $\tau_{110}$\cite{Hanson2007,Hanson2007b}.
A single rotation is enough to yield any state with $\theta\leq\theta_{\mathrm{max}}$,
where $\theta_{\mathrm{max}}=2\arcsin\left(|\tau_{110}|/\sqrt{\tau_{110}^{2}+\Delta\tilde{E}_{\mathrm{LC}}^{2}}\right)$.
For states with $\theta>\theta_{\mathrm{max}}$, an arbitrary $X$
axis rotation can be realized by applying three consecutive rotations\cite{Hanson2007b}.
If $\theta\leq\theta_{\mathrm{max}}$, the system acquires a finite
phase $\phi'$ while setting $\theta_{\mathrm{L}}$. Then, a second,
independent axis of control is given by raising the barriers, so that
$\tau_{110}\simeq0$. This yields a rotation around the $Z$ axis,
which can be used to set the desired value of $\phi_{\mathrm{L}}$
by letting the system evolve for a fraction of the period $\mathcal{T}_{\Delta}=\pi\hbar/|\Delta_{\mathrm{LC}}|$. 

As in Sec.~\ref{subsec:woGrads}, the quantum state transfer is initiated
by turning the ac voltage on. As before, $\tau_{110}$ needs to be
set to zero so that $\theta_{\mathrm{L}}$ does not vary during the
transfer process. In the presence of gradients, the dark state condition
leading to $\tau_{110}=0$ is given by
\begin{align*}
 & \sum_{\nu}J_{\nu}^{2}\left(\frac{V_{\text{ds}}^{\text{L}}}{\hbar\omega}\right)\left[\frac{1}{\delta_{020}+\Delta_{\mathrm{LC}}-\nu\hbar\omega}+\frac{1}{\delta_{200}+\Delta_{\mathrm{LC}}-\nu\hbar\omega}\right.\\
 & \left.+\frac{1}{\delta_{020}-\Delta_{\mathrm{LC}}-\nu\hbar\omega}+\frac{1}{\delta_{200}-\Delta_{\mathrm{LC}}-\nu\hbar\omega}\right]=0
\end{align*}
and similar conditions can be obtained for $\tau_{011}$ and $\tau_{\mathrm{LR,}2}^{n}$.
Furthermore, we will assume that $|\Delta_{\mathrm{LC}}-\Delta_{\mathrm{CR}}|\ll\omega$
so that the RWA approximation holds. 

The procedure for transferring the state from one edge to the other
depends on the gradient configuration. If the magnetic field gradients
are much smaller than the long-range transfer rate $\tau_{\mathrm{LR},1}^{n}$,
the state can be transferred directly without significant variation
in $\phi_{\mathrm{L}}$, following the same protocol as in the case
without magnetic gradients. Otherwise, two problems arise. First,
the finite gradient results in a change in $\phi_{\mathrm{L}}$ during
the transfer process. This can be circumvented by letting $\phi_{\mathrm{L}}$
evolve during a finite time after the transfer process is finished
in order to compensate the change in $\phi_{\mathrm{L}}$. The second
problem is the difference between the gradients in $\mathcal{Q}_{\mathrm{L}}$
and $\mathcal{Q}_{\mathrm{R}}$, which imposes a limit to the fidelity
of a transfer process (through the long-range channel $\tau_{\mathrm{LR,1}}^{n}$)
that can be estimated as\cite{Note1}
\begin{equation}
\mathcal{F}_{\mathrm{max}}\simeq1-\frac{|\Delta_{\mathrm{CR}}-\Delta_{\mathrm{LC}}|}{\sqrt{\left(\Delta_{\mathrm{CR}}-\Delta_{\mathrm{LC}}\right)^{2}+\left(2\tau_{\mathrm{LR,1}}^{n}\right){}^{2}}}.\label{eq:Fmax}
\end{equation}

\begin{figure}
\begin{centering}
\includegraphics[width=1\columnwidth]{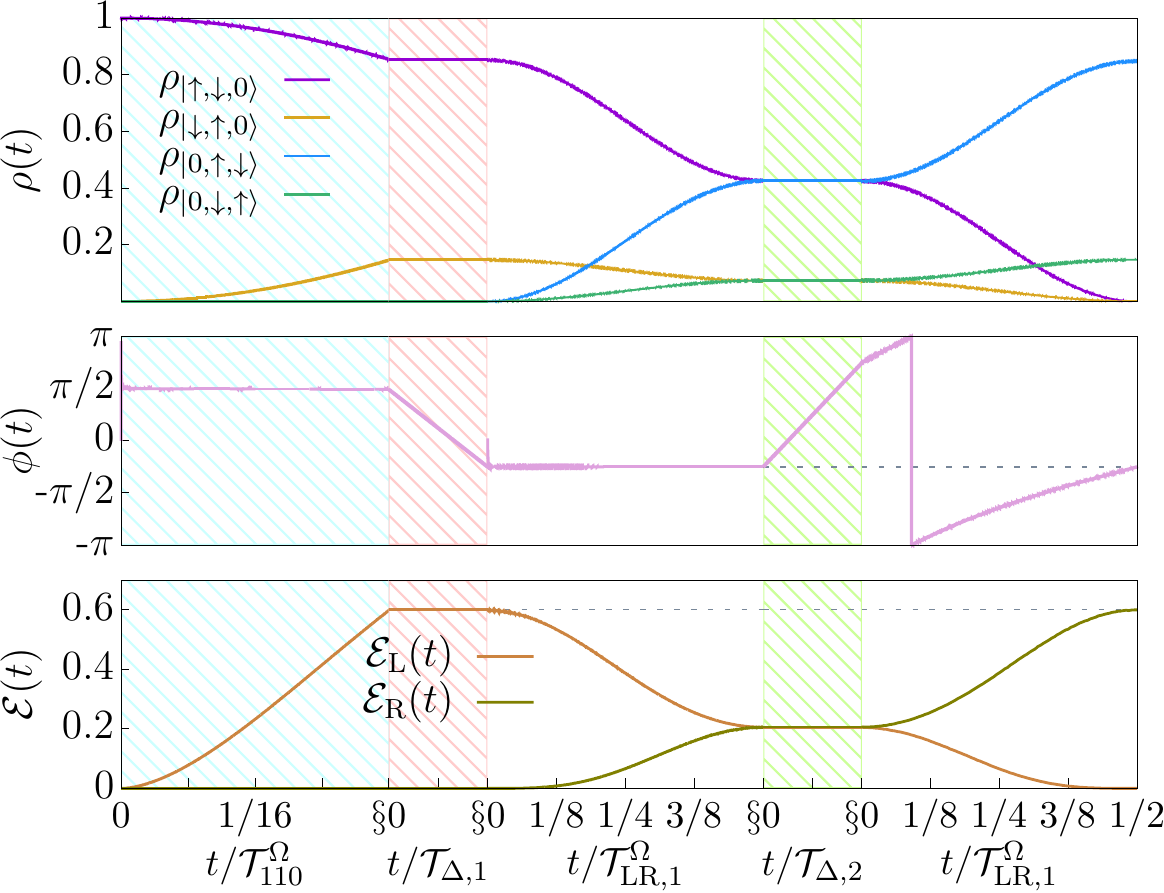}
\par\end{centering}
\caption{For $\Delta_{\mathrm{LC}}=-\Delta_{\mathrm{CR}}$: (top) Time evolution
of the population of the states. (center) Time evolution of $\phi(t)$.
The phase $\phi(t)$ is defined in $\mathcal{Q}_{\mathrm{L}}$ during
the manipulation process and in \emph{$\mathcal{Q}_{\mathrm{R}}$
}during the transfer process. (bottom) Time evolution of the entanglement
$\mathcal{E}(t)$ between the two spins. From left to right: in the
blue dashed area $\theta_{\mathrm{L}}$ is fixed to the desired value
of $\theta_{\mathrm{L}}=\pi/4$; in the red dashed area the polar
angle $\phi_{\mathrm{L}}=-\pi/4$ is set through the magnetic field
gradient while $\tau_{110}\simeq0$; in the first white area, the
state is transferred to a superposition with equal weight in $\mathcal{Q}_{\mathrm{L}}$
and\emph{ $\mathcal{Q}_{\mathrm{R}}$}; in the green area, the system
is left to evolve under the gradients, $\Delta_{\mathrm{LC}}$ and
$\Delta_{\mathrm{CR}}$, for a time $\mathcal{T}_{\Delta}/2=\pi/2|\Delta_{\mathrm{LC}}|$;
finally, in the second white area, the state is transferred from the
superposition between $\mathcal{Q}_{\mathrm{L}}$ and\emph{ $\mathcal{Q}_{\mathrm{R}}$}
to $\mathcal{Q_{\mathrm{R}}}$. Parameters: $\Delta_{\mathrm{LC}}=-\Delta_{\mathrm{CR}}=\tau_{\text{LR,1}}$.
$\tau_{\mathrm{LC}}=\tau_{\mathrm{CR}}=30$~$\mathrm{\mu eV}$ in
the white areas and the blue dashed areas and $\tau_{\mathrm{LC}}=\tau_{\mathrm{CR}}=0$
in the red and green dashed areas. , $\delta_{020}=\zeta_{020}=4.25$~$\text{meV}$,
$\delta_{101}=\zeta_{101}=2.28$~$\text{meV}$, $n=0$. In the dashed
areas the ac gate voltages are switched off. In the white areas: $\omega=0.5$~$\mathrm{meV}$
and $V_{\text{ac}}^{\text{L}}=V_{\mathrm{ac}}^{\mathrm{L}}=5.25\,\text{meV}$.
The gray dashed lines are a visual guide indicating the desired value
of $\phi$ and the entanglement of the initially prepared state.\label{fig:osRabi_SYM}}
\end{figure}

We consider first the simpler case of $\Delta_{\mathrm{LC}}=\Delta_{\mathrm{CR}}=\Delta$
(the \emph{linear }configuration). The estimated maximum fidelity,
Eq.~\ref{eq:Fmax} is 1 for this configuration. Hence, the only issue
with the presence of the gradients in this configuration is that the
phase $\phi_{\mathrm{L}}$ keeps evolving during the state transfer.
This can be circumvented by letting the phase evolve for a time 
\[
\mathcal{T}_{\mathrm{off}}=\left[N\mathcal{T}_{\Delta}-\mathcal{T}_{\mathrm{LR,1}}^{\Omega}/2\right],\,\,\,N\mathcal{2}\mathbb{Z}
\]
once the state has been transferred\textcolor{black}{, where 
\[
\mathcal{T}_{\mathrm{LR,1}}^{\Omega}=2\pi\hbar\left[\left(2\tau_{\mathrm{LR,1}}^{n=0}\right)^{2}+\left(\Delta_{\mathrm{LC}}-\Delta_{\mathrm{CR}}\right)^{2}\right]^{-1/2}.
\]
}is the Rabi period corresponding to $\tau_{\text{LR,1}}^{n=0}$,
written here \textcolor{black}{for arbitrary $\Delta_{\text{LC}},\Delta_{\text{CR}}$
for completeness.} $\mathcal{T}_{\mathrm{off}}$ does not depend on
the particular state being transferred and in that sense the process
is still universal. The process is illustrated in \textcolor{black}{Fig.~\ref{fig:osRabi_LIN}.
Initially, we set $\tau_{\mathrm{CR}}=0$ since the levels are not
detuned; in the blue dashed part, $\theta_{\mathrm{L}}$ is fixed
to its desired value of $\pi/4$ through $\tau_{110}$. Note that
$\tau_{110}\gg\Delta_{\mathrm{LC}}$, and therefore $\phi'\simeq\pi/2$
when $\theta_{\mathrm{L}}$ reaches $\pi/4$. In the red dashed area,
we set $\tau_{110}=0$ and the phase evolves from $\phi'$ to $\phi_{\mathrm{L}}$
(marked by a gray dashed line). Then, the two barriers are lowered
and the transfer process is carried out for a time $\mathcal{T}_{\mathrm{LR,1}}^{\Omega}/2$.
Finally, in the green dashed area, the barriers are raised again and
the phase $\phi_{\mathrm{L}}$ is left to evolve for a time $\mathcal{T}_{\mathrm{off}}$
until the desired value $\phi_{\mathrm{L}}$ is reached.}

If $\Delta_{\mathrm{LC}}\neq\Delta_{\mathrm{CR}}$, the maximum fidelity,
Eq.~\ref{eq:Fmax}, cannot reach 1 for a transfer operation in a
single step. Hence, there are two options to transfer the state: (i)
if the difference between the gradients is much smaller than the long-range
transition rates, $\left|\Delta_{\mathrm{LC}}-\Delta_{\mathrm{CR}}\right|\ll\left|\tau_{\mathrm{LR,1}}^{n}\right|$,
and so $\mathcal{F}_{\mathrm{max}}\simeq1$ for a single transfer
operation; (ii) a combination of operations in one of the two level
systems $\mathcal{Q}_{\mathrm{L}}$ and $\mathcal{Q}_{\mathrm{R}}$
and transfer operations through $\tau_{\mathrm{LR,1}}^{n}$ is used
to ensure an ideal 100\% fidelity at the cost of longer transfer times.
The latter case is discussed in detail in the Supplementary Information
and a universal transfer process with 100\% fidelity for any $\{\Delta_{\mathrm{LC}},\Delta_{\mathrm{CR}}\}$
configuration is proposed. 

Here, we consider for example $\Delta_{\mathrm{LC}}=-\Delta_{\mathrm{CR}}$
in Fig.~\ref{fig:osRabi_SYM} (the \emph{symmetric }configuration).
This configuration has the particularity that $\phi_{\mathrm{L}}$
is not modified during a single transfer process, as can be seen in
the first white section of Fig.~\ref{fig:osRabi_SYM}~(center).
On the other hand, the maximum fidelity of a single transfer operation,
as given in Eq.~\ref{eq:Fmax}, is minimal for this configuration.
To transfer the state with 100\% ideal fidelity, a sequence consisting
of (i) transfer from $\mathcal{Q_{\mathrm{L}}}$ to a superposition
of the desired state with equal weight in $\mathcal{Q}_{\mathrm{L}}$
and $\mathcal{Q}_{\mathrm{R}}$, (ii) evolution under the gradients,
$\Delta_{\mathrm{LC}}$ and $\Delta_{\mathrm{CR}}$, for a time $\mathcal{T}_{\Delta}/2=\pi\hbar/\left(2|\Delta_{\mathrm{LC}}|\right)$
and (iii) another transfer process as in (i) from the superposition
between $\mathcal{Q}_{\mathrm{L}}$ and $\mathcal{Q}_{\mathrm{R}}$
to $\mathcal{Q}_{\mathrm{R}}$, can be used to transfer the state
with maximum fidelity. The operations (i)-(iii) correspond to the
first white area, the green dashed area and the second white area
of Fig.~\ref{fig:osRabi_SYM}, respectively. Each of the transfer
operations, (i) and (iii) is carried out \textcolor{black}{for a time
$\mathcal{T}_{\mathrm{LR,1}}^{\Omega}/2$.}

If the magnetic field gradients could be be switched off rapidly enough
during operation, the transfer protocol could be performed in the
simpler manner of Section~\ref{subsec:woGrads} (i.e. independent
of the gradient configuration). This has been recently shown to be
possible in reasonable operation times\cite{1804.02893}. 

\subsection{Relaxation and decoherence\label{subsec:Relaxation-and-decoherence}}

In this section we will discuss the effect of relaxation and decoherence
on the protocol. There are several possible sources of decoherence
in these systems, but the most important is the coupling to charge
noise. We will discuss charge noise first and later on we will consider
other sources of decoherence. For charge noise we will search for
optimal operation points (\emph{sweetspots}) under which the coupling
to charge noise is minimized.

\subsubsection{Charge noise}

In order to estimate the effect of charge noise we consider that the
system is coupled to a bath consisting of a set of independent harmonic
oscillators. The Hamiltonian for the system and bath is given by $H=H_{\mathrm{S}}(t)+H_{\mathrm{B}}+H_{\mathrm{SB}}$,
where $H_{\mathrm{S}}(t)$ is the Hamiltonian for the system as given
by Eq.~\ref{eq:EffH} and
\begin{align}
H_{\mathrm{B}} & =\sum_{i,n}\hbar\omega_{n}\hat{b}_{i,n}^{\dagger}\hat{b}_{i,n}\label{eq:HB}\\
H_{\mathrm{SB}} & =\sum_{i}X_{i}\xi_{i}=\sum_{i,n}g_{n}X_{i}(\hat{b}_{i,n}^{\dagger}+\hat{b}_{i,n})\label{eq:HSB}
\end{align}

$\{X_{i}\}$ is the set of system operators coupled to the bath. In
our case we consider only charge noise, corresponding to $X_{i}=\hat{c}_{i}^{\dagger}\hat{c}_{i}$.
We assume that all oscillators are equal and independent. For the
bath coordinates $\{\xi_{i}\}$ this requires that the symmetrically
ordered autocorrelation function satisfies $(1/2)\mathcal{h}\left\{ \xi_{i}(\tau),\xi_{i}(0)\right\} \mathcal{i}=2\delta(t-t')\delta_{ij}$.
Current noise has a small effect in quantum dot-based quantum information
devices\cite{Russ2017} and we will not consider it here. The bath
is characterized by the spectral density, $\mathcal{J}(\omega)=\pi\sum_{n}|g_{n}|^{2}\delta(\omega-\omega_{n})$,
and by $\mathcal{S}(\omega)=\mathcal{J}(\omega)\mathrm{coth}(\beta\hbar\omega/2)$,
the Fourier transform of the symmetrically ordered equilibrium autocorrelation
function.

The system under the presence of charge noise can be studied under
a Bloch-Redfield type master equation\cite{Redfield1957,Kohler2006,Qi2017}.
For the $1/f$ noise typically considered in quantum dot systems,
the validity of the Markovian approximation inherent in a master equation
approach is only warranted for weak coupling. At this level of approximation,
$1/f$ noise can be considered by taking $\mathcal{J}\left(\omega\right)\simeq\text{constant}$,
which gives $\mathcal{S}\left(\omega\right)\sim1/\omega$ for $\beta\hbar\omega\ll1$.
Since $\mathcal{S}\left(\omega\right)$ diverges for low frequencies,
we regularize it below a certain cutoff frequency $\omega_{\text{IR}}$
as 

\begin{equation}
\mathcal{S}\left(\omega\right)=\begin{cases}
\mathcal{S}_{0} & \omega\leq\omega_{\text{IR}}\\
\mathcal{S}_{0}\frac{\tanh\left(\beta\hbar\omega_{\text{IR}}/2\right)}{\tanh\left(\beta\hbar\omega/2\right)} & \omega>\omega_{\text{IR}}
\end{cases}\label{eq:Somega}
\end{equation}

The parameter $\mathcal{S}_{0}$ determines the dephasing time, and
thus provides a natural parameter to characterize the noise intensity.
We will consider the effect of noise in the protocol both for the
process of manipulation and transfer. 

Charge noise comes from fluctuations on the energy levels. During
the  manipulation process, it modifies the renormalized splitting
$\Delta\tilde{E}_{\mathrm{LC}}$, given by Eq.~\ref{eq:renorm-energy-split}
(with $V_{\mathrm{L}}^{\mathrm{ac}}=V_{\mathrm{R}}^{\mathrm{ac}}=0$),
and the transition rate $\tau_{110}$, given by Eq.~\ref{eq:tau110-no-ac},
associated to the exchange interaction. The system is effectively
subjected to a single noise source $\xi_{\mathrm{LC}}=\xi_{\mathrm{C}}-\xi_{\mathrm{L}}$.
Defining $\tau_{110}^{(1)}\equiv\left.\partial_{\epsilon_{\mathrm{L}}}\tau_{110}\right|_{\xi_{\mathrm{LC}}=0}$,
and $\Delta\tilde{E}_{\mathrm{LC}}^{(1)}=\left.\partial_{\epsilon_{\mathrm{L}}}\Delta\tilde{E}_{\mathrm{LC}}\right|_{\xi_{\mathrm{LC}}=0}$,
under the conditions $\tau_{110}^{(1)}=0$ and $\Delta\tilde{E}_{\mathrm{LC}}^{(1)}=0$,
the system is unaffected by charge noise, yielding the previously
mentioned sweetspot\cite{Fei2015,Martins2016}. The non-linear terms
are only predominant at the sweetspot, but their treatment is complex\cite{Makhlin2004}
and we will not consider them any further. The condition $\tau_{110}^{(1)}=0$
implies that
\begin{equation}
\epsilon_{\mathrm{C}}-\epsilon_{\mathrm{L}}=\frac{U_{\mathrm{LL}}-U_{\mathrm{CC}}}{2}.\label{eq:noisecon1}
\end{equation}

At the sweetspot, the transition rate $\tau_{110}$ is given by 
\begin{align*}
\tau_{110} & =-\tau_{\mathrm{LC}}^{2}\left(\frac{1}{\delta_{\mathrm{ss}}+\Delta_{\mathrm{LC}}}+\frac{1}{\delta_{\mathrm{ss}}-\Delta_{\mathrm{LC}}}\right)
\end{align*}
where $\delta_{\mathrm{ss}}=(U_{\mathrm{LL}}+U_{\mathrm{CC}})/2-U_{\mathrm{LC}}$.
Under the sweetspot condition, Eq.~\ref{eq:noisecon1}, $\Delta\tilde{E}_{\mathrm{LC}}^{(1)}=0$
as long as 
\[
\tau_{\mathrm{CR}}^{2}\left(\frac{1}{\delta_{101}+\Delta_{\mathrm{CR}}}-\frac{1}{\delta_{101}-\Delta_{\mathrm{CR}}}\right)=0
\]
which is satisfied for $\tau_{\text{CR}}=0$ or $\Delta_{\text{CR}}=0$.
The sweetspot for manipulation in $\mathcal{Q}_{\mathrm{R}}$ can
be obtained in the same manner.

During the transfer process, the system couples to charge noise through
several processes: 
\begin{enumerate}
\item Direct coupling to charge noise through the energy levels of the quantum
dots ($\epsilon_{\mathrm{L}},\epsilon_{\mathrm{C}},\epsilon_{\mathrm{R}}$).
The system-bath interaction for this process is given by
\begin{equation}
H_{\mathrm{SB,dir}}=\sum_{i}\xi_{i}\hat{c}_{i}^{\dagger}\hat{c}_{i}\label{eq:HSBdir}
\end{equation}
\item Through the energy-dependence of the long-range amplitude $\tau_{\mathrm{LR,1}}$.
The related relaxation and dephasing rates are proportional to the
first derivative of $\tau_{\mathrm{LR,1}}$ with respect to the gate
energy $\epsilon_{\mathrm{L}}$ (or $\epsilon_{\mathrm{R}}$), denoted
by $\tau_{\mathrm{LR,1}}^{(1)}$.
\item Because charge noise disrupts the dark state condition and results
in non-zero values for $\tau_{110},\tau_{011}$ and $\tau_{\mathrm{LR,2}}$.
The related relaxation and dephasing rates are proportional to the
first derivatives $\tau_{110}^{(1)},\tau_{011}^{(1)}$ and $\tau_{\mathrm{LR,2}}^{(1)}$,
respectively.
\end{enumerate}
As a result of these three processes, the system is effectively coupled
to three noise sources, $\xi_{\mathrm{LC}},\xi_{\mathrm{CR}},\xi_{\mathrm{LR}}$,
where $\xi_{ij}=\xi_{j}-\xi_{i}$. 

The direct coupling to noise (process 1) is by far the dominant source
of decoherence and relaxation. This can be seen by inspection of the
decay rates between the different eigenstates, which are obtained
analytically in the Supplementary Information. The relaxation rate
due to direct coupling to noise by the Hamiltonian Eq.~\ref{eq:HSBdir}
is obtained as
\begin{equation}
\Gamma_{\mathrm{dir}}=\mathcal{S}(\Lambda)\left(\frac{1}{2}\sin\Upsilon\right)^{2},\label{eq:gamma-dir}
\end{equation}
where $\Lambda=\sqrt{(\Delta_{\mathrm{LC}}-\Delta_{\mathrm{CR}})^{2}+4\tau_{\mathrm{LR,1}}^{2}}$
and $\Upsilon$ is given by
\begin{align}
\tau_{\mathrm{LR,1}}\tan\left(\frac{\Upsilon}{2}\right)=\frac{1}{2}\left(\Delta_{\mathrm{LC}}-\Delta_{\mathrm{CR}}+\Lambda\right).\label{eq:upsilon}
\end{align}

This can be compared, for instance, with the relaxation rate due to
the coupling to noise via the energy-dependence of $\tau_{\mathrm{LR,1}}$
(process 2), given by
\[
\Gamma_{\mathrm{LR,1}}=\mathcal{S}(\Lambda)\left(\tau_{\mathrm{LR,1}}^{(1)}\cos\Upsilon\right)^{2},
\]

Since, $\Gamma_{\mathrm{dir}}/\Gamma_{\mathrm{LR,1}}\sim|\tau_{\mathrm{LC}}\tau_{\mathrm{CR}}|^{-2}$,
$\Gamma_{\mathrm{dir}}$ is the largest source of decoherence by a
factor $|\tau_{\mathrm{LC}}\tau_{\mathrm{CR}}|^{-2}$. Furthermore,
for the coupling to noise via the energy-dependence of $\tau_{\mathrm{LR,1}}$,
a noise sweetspot can be found, 
\[
\left.\partial_{\epsilon_{\mathrm{L}}}\tau_{\mathrm{LR,1}}\right|_{\xi_{\mathrm{LC}}=0}=\left.\partial_{\epsilon_{\mathrm{R}}}\tau_{\mathrm{LR,1}}\right|_{\xi_{\mathrm{CR}}=0}=0.
\]

Under the RWA approximation, this can be written as

\begin{widetext}

\begin{align}
\sum_{\nu}J_{\nu}\left(\frac{V_{\mathrm{L}}^{\mathrm{ac}}}{\hbar\omega}\right)J_{\nu-n}\left(\frac{V_{\mathrm{R}}^{\mathrm{ac}}}{\hbar\omega}\right)\left[\frac{1}{\left(\delta{}_{020}+\Delta_{\mathrm{LC}}-\nu\hbar\omega\right)^{2}}+\frac{1}{\left(\zeta_{101}+\Delta_{\mathrm{LC}}+\nu\hbar\omega\right)^{2}}\right] & =0,\label{eq:sweetspot-tLR1LC}\\
\sum_{\nu}J_{\nu}\left(\frac{V_{\mathrm{L}}^{\mathrm{ac}}}{\hbar\omega}\right)J_{\nu-n}\left(\frac{V_{\mathrm{R}}^{\mathrm{ac}}}{\hbar\omega}\right)\left[\frac{1}{\left(\zeta_{020}+\Delta_{\mathrm{CR}}-(\nu-n)\hbar\omega\right)^{2}}+\frac{1}{\left(\delta_{101}+\Delta_{\mathrm{CR}}+(\nu-n)\hbar\omega\right)^{2}}\right] & =0.\label{eq:sweetspot-tLR1CR}
\end{align}

\end{widetext}

\begin{figure}
\includegraphics[width=1\columnwidth]{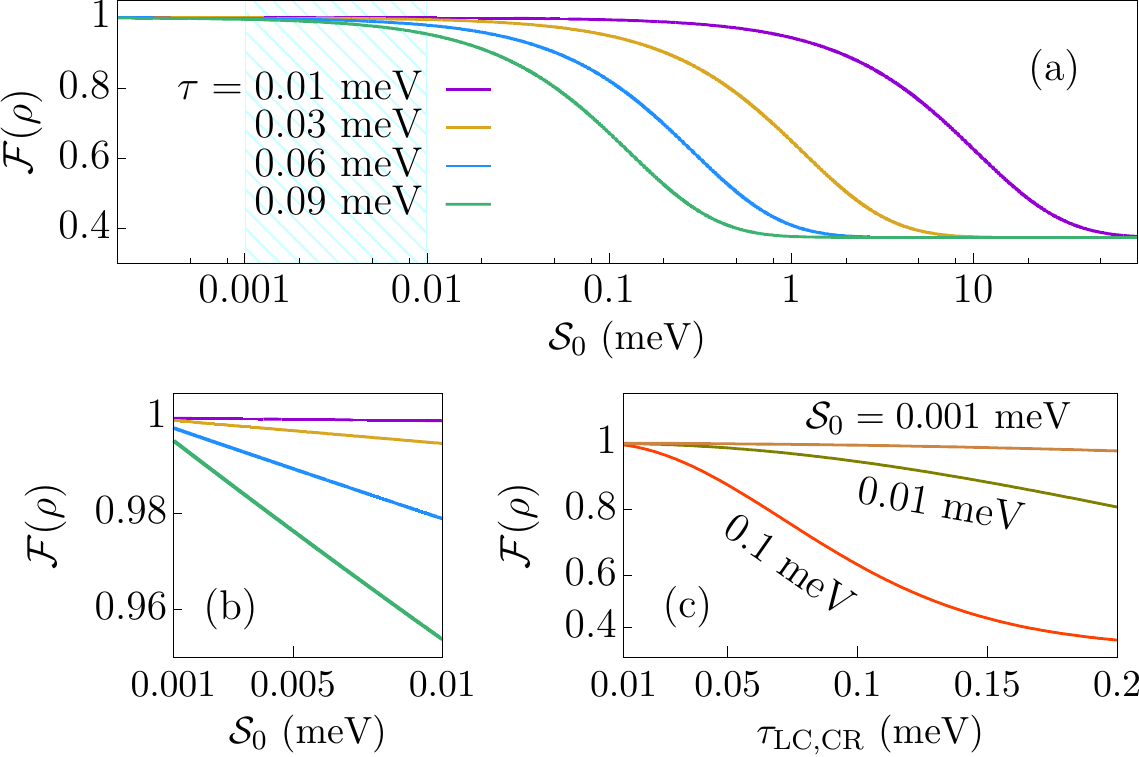}

\caption{(a) Fidelity of the transfer process as a function of the effective
noise intensity $\mathcal{S}_{0}$ for a state prepared with $\theta_{\mathrm{L}}=\pi/4$
and $\phi_{\mathrm{L}}=-\pi/4$ calculated with the Bloch-Redfield
master equation, Eq.~\ref{eq:masterEq} using the effective Hamiltonian,
Eq.~\ref{eq:EffH} in the RWA approximation. The different curves
correspond to $\tau=\tau_{\text{LC}}=\tau_{\text{CR}}=10$~(purple),
$30$~(yellow),~$60$~(blue), and~$90$ (green)~$\mu$eV. In
(b) we have highlighted the range $\mathcal{S}_{0}=1-10$~$\mu\text{eV}$.
(c) Fidelity as a function of $\tau_{\text{LC}}=\tau_{\text{CR}}$
for $\mathcal{S}_{0}=0.1$~(red), $0.01$~(green) and $0.001$~(orange)The
parameters are as in Fig.~\ref{fig:osRabi_LIN} with $T=1$~K and
infrared cutoff 1~neV. \label{fig:FidelNoise}}
\end{figure}

Contrary to the sweetspots for $\tau_{110}$ and $\tau_{011}$ in
the manipulation process, the sweetspot corresponding to the conditions
of Eqs~\ref{eq:sweetspot-tLR1LC} and~\ref{eq:sweetspot-tLR1CR},
is induced by the ac-voltage. The sweetspot for $\tau_{110}$ in Eq.~\ref{eq:noisecon1}
appears because the dependence on the energies $\epsilon_{i}$ coming
from virtual transitions to the $(\up\dw,0,0)$ and $(0,\up\dw,0)$
states compensate each other. In the sweetspots of Eqs.~\ref{eq:sweetspot-tLR1LC}
and~\ref{eq:sweetspot-tLR1CR}, however, it is the dependence on
the gate energies coming from different ac-induced sidebands that
compensate each other. 

Finally, the sweetspot condition for deviations from the dark state
condition is $\tau_{110}^{(1)}=0$ and $\tau_{011}^{(1)}=\left.\partial_{\epsilon_{\mathrm{R}}}\tau_{011}\right|_{\xi_{\mathrm{CR}}=0}=0$,
yielding the same conditions on the gate energies, Eq.~\ref{eq:noisecon1},
as in the manipulation process. 

Since direct coupling is the dominant contribution from charge noise,
we discuss it in detail. In Sec.~V of the Supplementary Information,
we write explicitly the Bloch-Redfield operator $Q_{\text{LR}}^{\text{dir}}$
that results from direct coupling to noise (see Eq.~\ref{eq:Qj}).
We see that there are two contributions. The first is proportional
to $\mathcal{S}\left(\Lambda\right)\sin\Upsilon$ and is the one responsible
for relaxation, as can be seen from the expression for the relaxation
rate due to direct coupling, Eq.~\ref{eq:gamma-dir}. The second
contribution is proportional to $\mathcal{S}\left(0\right)\cos\Upsilon$
and is the one responsible for dephasing. Since $\mathcal{S}\left(0\right)\gg\mathcal{S}\left(\Lambda\right)$,
this is also the most important of the two. However, it vanishes for
$\cos\Upsilon=0$, that is, for for $\Upsilon=(2n+1)\pi/2$, $n\mathcal{2}\mathbb{Z}$.
From Eq.~\ref{eq:upsilon} we see that this corresponds to $\Delta_{\text{LC}}=\Delta_{\text{CR}}$,
which includes both the case in which the gradients are negligible
or can be turned off, and the linear configuration discussed in Sec.~\ref{subsec:withGrads}.
Hence, this configuration provides the best protection against charge
noise. 

In Fig.~\ref{fig:FidelNoise} (a) we have plotted the fidelity as
a function of $\mathcal{S}_{0}$. We perform the calculations for
the case $\Delta_{\text{LC}}=\Delta_{\text{CR}}$ under the dark state
condition. We employ values of $\tau_{\text{LR,1}}$ compatible with
the values for the exchange interaction in Ref.~\cite{Dial2013},
corresponding to $\tau_{\text{LC}}=\tau_{\text{CR}}=10$~(purple),
$30$~(yellow),~$60$~(blue), and~$90$ (green)~$\mu$eV. In
Fig.~\ref{fig:FidelNoise} (b) we have plotted the fidelity in the
realistic range $\mathcal{S}_{0}\sim1-10\mu\text{eV}$\cite{Dial2013,Wu2014,Qi2017}.
For $\tau_{\text{LC}}=\tau_{\text{CR}}=10\,\mu\text{eV}$ we obtain
a fidelity of 99.99\% for $\mathcal{S}_{0}=1\mu eV$ and of 99.93\%
for $\mathcal{S}_{0}=10\mu eV$ . Fig.~\ref{fig:FidelNoise} (c)
we have plotted the fidelity as a function of $\tau_{\text{LC}}=\tau_{\text{CR}}$
for the different noise intensities $\mathcal{S}_{0}=0.1$~(red),
$0.01$~(green) and $0.001$~(orange). By our results we observe
that in this realistic range, decreasing $\tau_{\text{LC}},\tau_{\text{CR}}$
increases the total fidelity when considering only charge noise. This
can be explained in the following way. For $\Delta_{\text{LC}}=\Delta_{\text{CR}}$,
the dominating dephasing process comes from the energy-dependence
of the long-range amplitude $\tau_{\mathrm{LR,1}}$. In that case,
increasing $\tau_{\text{LC}}$ and $\tau_{\text{CR}}$ to reduce the
transfer time also increases the dominant dephasing rate. On the other
hand, decreasing $\tau_{\text{LC}},\tau_{\text{CR}}$ reduces $\Lambda$,
and in turn increases $\mathcal{S}\left(\Lambda\right)$, but this
effect is of lesser importance. 

\subsubsection{Other sources}

Although charge noise is the most significant source of decoherence,
magnetic noise caused by the hyperfine coupling and fluctuations in
the gradients also detracts from the fidelity. As a result, the spin
nuclear bath induces a time-scale, $\mathcal{T}_{2}^{\left(\text{HF}\right)}$,
under which the state transfer can be realized with minimal fidelity
losses. As shown in Fig.~\ref{fig:FidelNoise} (c), reducing $\tau_{\text{LC}},\tau_{\text{CR}}$
is beneficial to limit the effect of charge noise. If as a result
of reducing $\tau_{\text{LC}},\tau_{\text{CR}}$, the transfer time
$\mathcal{T}_{\text{LR},1}$ is increased above $\mathcal{T}_{2}^{\left(\text{HF}\right)}$,
the hyperfine-induced dephasing disrupts the transferred state. Furthermore,
relaxation leads to leakage to the states $\left\{ \ket{\up,\up,0},\ket{\dw,\dw,0},\ket{0,\up,\up},\ket{0,\dw,\dw}\right\} $,
which affects the entanglement $\mathcal{E}(\rho)$ through the concurrence
$\mathcal{C}$\cite{Note1}. The effect of the hyperfine interaction
can be overcome by employing isotope purification in Silicon qubits. 

Other effects that may detract from the fidelity are finite ramping
times\cite{Li2017}, tunnel noise\cite{Russ2016,Bello2017}, spin-dependent
tunneling rates\cite{Danon2009,Schreiber2011} and multiple valley
states in Silicon\cite{Yang2013,Veldhorst2015}, although most can
be reduced by other means\cite{Li2017}.

\section{Discussion\label{sec:Discussion}}

In summary, we propose a fully tunable two-level system in a double
quantum dot contained in one edge of a triple quantum dot structure.
By means of ac gate voltages a prepared quantum state in one edge
can be transferred to another two-level system defined at the other
edge of the TQD by means of photoassisted virtual transitions. The
ac voltages fix the prepared state by blocking virtual transitions
that modify the desired state and suppress undesired transfer channels
via the formation of dark states. In order to measure the information
transfer between the two two-level systems we have calculated the
time evolution of the states occupations, the phase and their entanglement.
The set-up is limited by charge and magnetic noise; the former is
induced by random variations in the gate energies and the second by
the hyperfine interaction and fluctuations in the gradients. The effect
of charge noise can be alleviated by working at the noise sweetspots,
where the system is first-order insensitive to charge noise. In that
regard, we have shown how the interference between sidebands can induce
a sweetspot that does not exist without ac voltages. The latter essentially
imposes a time-scale under which the operation can be realized effectively.
We show that the protocol has a fidelity $>99\%$ for realistic values
of the charge noise intensity and the tunnel barriers.The efficiency
of the protocol for quantum state transfer could be improved by considering
Si quantum dots where spin flip induced by hyperfine interaction can
be strongly reduced through isotope purification. If the transfer
times are faster than the decoherence times, the procedure can be
generalized to longer quantum dot arrays by using the general state
transfer protocol for arbitrary gradient configurations sequentially.
Furthermore, the protocol can be implemented experimentally with available
technologies, which are no different than those employed to manipulate
the exchange interaction in quantum dot-based qubits. Operating in
the sweetspots reduces considerably the difficulty in finding the
dark state condition required to suppress unwanted processes, leaving
the possibility within experimental bounds. We also expect that the
technique of dark state formation with ac driving can be employed
in the future in other setups to suppress or mitigate processes detrimental
to the fidelity of quantum gates or for the possibility of inducing
dynamical sweetspots. 

\section*{Methods}

The time-evolution of the density matrix under the assumptions of
weak coupling and Markovianity is given by the Bloch-Redfield master
equation~
\begin{align}
\dot{\rho}(t) & =-i\hbar^{-1}\left[H_{\mathrm{S}},\rho(t)\right]\nonumber \\
 & -\sum_{i}\left[X_{i},\left[Q_{i},\rho(t)\right]\right]-\sum_{i}\left[X_{i},\left\{ R_{i},\rho(t)\right\} \right]\label{eq:masterEq}
\end{align}
where $\left\{ \,,\,\right\} $ indicates the anti-commutator and
\begin{equation}
Q_{i}=\frac{1}{\pi}\int_{0}^{\infty}d\tau\int_{0}^{\infty}d\omega\mathcal{S}(\omega)\tilde{X}_{i}(\tau,0)\cos(\omega\tau)\label{eq:Qj}
\end{equation}
\begin{equation}
R_{i}=\frac{1}{i\pi}\int_{0}^{\infty}d\tau\int_{0}^{\infty}d\omega\mathcal{J}(\omega)\tilde{X}_{i}(\tau,0)\sin(\omega\tau)\label{eq:Rj}
\end{equation}

We define the propagated system operators as $\tilde{X}_{i}(\tau,\tau')=\mathcal{U}^{\dagger}(\tau,\tau')X_{i}\mathcal{U}(\tau,\tau')$.
Apart from $X_{i}=\hat{c}_{i}^{\dagger}\hat{c}_{i}$, we have to consider
the coupling between system and bath through the virtual tunneling
processes, which depend on the energy differences between the states. 

For sections~\ref{subsec:woGrads} and \ref{subsec:withGrads}, results
are obtained without any source of decoherence. Then, Eq.~\ref{eq:masterEq}
reduces to $\dot{\rho}(t)=-i\hbar^{-1}\left[H_{\mathrm{S}},\rho(t)\right]$.
In these sections, the results are obtained with $H_{\text{S}}$ corresponding
to the full Hamiltonian of Eq.~\ref{eq:Hrot}. For section~\ref{subsec:Relaxation-and-decoherence},
the calculations are performed with the master equation, Eq.~\ref{eq:masterEq}.
The Hamiltonian $H_{\text{\text{\ensuremath{S}}}}$ for this section
is the effective Hamiltonian of Eq.~\ref{eq:EffH} in the time-independent
RWA. The incoherent terms appearing in Eq.~\ref{eq:masterEq} are
discussed the Supplementary Information.

\section*{Contributions}

J. Pic\'o-Cort\'es and F. Gallego-Marcos have contributed in developing
the theoretical model and the numerical calculations. G. Platero has
contributed in developing the theoretical model and has supervised
the work.
\begin{acknowledgments}
We acknowledge Rafael S\'anchez, Stefan Ludwig and Sigmund Kohler for
enlightening discussions and a critical reading of the manuscript.
This work was supported by the Spanish Ministry of Economy and Competitiveness
(MICINN) via Grants No. MAT2014-58241-P and MAT-2017-86717-P, 
the Youth Employment Initiative together with the Community of Madrid,
Exp. PEJ15/IND/AI-0444 and the Deutsche Forschungsgemeinschaft via SFB 1277-B4.
\end{acknowledgments}


\begin{thebibliography}{49}%
	\makeatletter
	\providecommand \@ifxundefined [1]{%
		\@ifx{#1\undefined}
	}%
	\providecommand \@ifnum [1]{%
		\ifnum #1\expandafter \@firstoftwo
		\else \expandafter \@secondoftwo
		\fi
	}%
	\providecommand \@ifx [1]{%
		\ifx #1\expandafter \@firstoftwo
		\else \expandafter \@secondoftwo
		\fi
	}%
	\providecommand \natexlab [1]{#1}%
	\providecommand \enquote  [1]{``#1''}%
	\providecommand \bibnamefont  [1]{#1}%
	\providecommand \bibfnamefont [1]{#1}%
	\providecommand \citenamefont [1]{#1}%
	\providecommand \href@noop [0]{\@secondoftwo}%
	\providecommand \href [0]{\begingroup \@sanitize@url \@href}%
	\providecommand \@href[1]{\@@startlink{#1}\@@href}%
	\providecommand \@@href[1]{\endgroup#1\@@endlink}%
	\providecommand \@sanitize@url [0]{\catcode `\\12\catcode `\$12\catcode
		`\&12\catcode `\#12\catcode `\^12\catcode `\_12\catcode `\%12\relax}%
	\providecommand \@@startlink[1]{}%
	\providecommand \@@endlink[0]{}%
	\providecommand \url  [0]{\begingroup\@sanitize@url \@url }%
	\providecommand \@url [1]{\endgroup\@href {#1}{\urlprefix }}%
	\providecommand \urlprefix  [0]{URL }%
	\providecommand \Eprint [0]{\href }%
	\providecommand \doibase [0]{http://dx.doi.org/}%
	\providecommand \selectlanguage [0]{\@gobble}%
	\providecommand \bibinfo  [0]{\@secondoftwo}%
	\providecommand \bibfield  [0]{\@secondoftwo}%
	\providecommand \translation [1]{[#1]}%
	\providecommand \BibitemOpen [0]{}%
	\providecommand \bibitemStop [0]{}%
	\providecommand \bibitemNoStop [0]{.\EOS\space}%
	\providecommand \EOS [0]{\spacefactor3000\relax}%
	\providecommand \BibitemShut  [1]{\csname bibitem#1\endcsname}%
	\let\auto@bib@innerbib\@empty
	%</preamble>
	\bibitem [{\citenamefont {Cirac}\ \emph {et~al.}(1997)\citenamefont {Cirac},
		\citenamefont {Zoller}, \citenamefont {Kimble},\ and\ \citenamefont
		{Mabuchi}}]{Cirac1997}%
	\BibitemOpen
	\bibfield  {author} {\bibinfo {author} {\bibfnamefont {J.~I.}\ \bibnamefont
			{Cirac}}, \bibinfo {author} {\bibfnamefont {P.}~\bibnamefont {Zoller}},
		\bibinfo {author} {\bibfnamefont {H.~J.}\ \bibnamefont {Kimble}}, \ and\
		\bibinfo {author} {\bibfnamefont {H.}~\bibnamefont {Mabuchi}},\ }\href
	{\doibase 10.1103/PhysRevLett.78.3221} {\bibfield  {journal} {\bibinfo
			{journal} {Phys. Rev. Lett.}\ }\textbf {\bibinfo {volume} {78}},\ \bibinfo
		{pages} {3221} (\bibinfo {year} {1997})}\BibitemShut {NoStop}%
	\bibitem [{\citenamefont {Vermersch}\ \emph {et~al.}(2017)\citenamefont
		{Vermersch}, \citenamefont {Guimond}, \citenamefont {Pichler},\ and\
		\citenamefont {Zoller}}]{Vermersch2017}%
	\BibitemOpen
	\bibfield  {author} {\bibinfo {author} {\bibfnamefont {B.}~\bibnamefont
			{Vermersch}}, \bibinfo {author} {\bibfnamefont {P.-O.}\ \bibnamefont
			{Guimond}}, \bibinfo {author} {\bibfnamefont {H.}~\bibnamefont {Pichler}}, \
		and\ \bibinfo {author} {\bibfnamefont {P.}~\bibnamefont {Zoller}},\ }\href
	{\doibase 10.1103/PhysRevLett.118.133601} {\bibfield  {journal} {\bibinfo
			{journal} {Phys. Rev. Lett.}\ }\textbf {\bibinfo {volume} {118}},\ \bibinfo
		{pages} {133601} (\bibinfo {year} {2017})}\BibitemShut {NoStop}%
	\bibitem [{\citenamefont {McNeil}\ \emph {et~al.}(2011)\citenamefont {McNeil},
		\citenamefont {Kataoka}, \citenamefont {Ford}, \citenamefont {Barnes},
		\citenamefont {Anderson}, \citenamefont {Jones}, \citenamefont {Farrer},\
		and\ \citenamefont {Ritchie}}]{McNeil2011}%
	\BibitemOpen
	\bibfield  {author} {\bibinfo {author} {\bibfnamefont {R.~P.~G.}\
			\bibnamefont {McNeil}}, \bibinfo {author} {\bibfnamefont {M.}~\bibnamefont
			{Kataoka}}, \bibinfo {author} {\bibfnamefont {C.~J.~B.}\ \bibnamefont
			{Ford}}, \bibinfo {author} {\bibfnamefont {C.~H.~W.}\ \bibnamefont {Barnes}},
		\bibinfo {author} {\bibfnamefont {D.}~\bibnamefont {Anderson}}, \bibinfo
		{author} {\bibfnamefont {G.~A.~C.}\ \bibnamefont {Jones}}, \bibinfo {author}
		{\bibfnamefont {I.}~\bibnamefont {Farrer}}, \ and\ \bibinfo {author}
		{\bibfnamefont {D.~A.}\ \bibnamefont {Ritchie}},\ }\href {\doibase
		10.1038/nature10444} {\bibfield  {journal} {\bibinfo  {journal} {Nature}\
		}\textbf {\bibinfo {volume} {477}},\ \bibinfo {pages} {439} (\bibinfo {year}
		{2011})}\BibitemShut {NoStop}%
	\bibitem [{\citenamefont {He}\ \emph {et~al.}(2017)\citenamefont {He},
		\citenamefont {He}, \citenamefont {Wei}, \citenamefont {Jiang}, \citenamefont
		{Chen}, \citenamefont {Lu}, \citenamefont {Pan}, \citenamefont {Schneider},
		\citenamefont {Kamp},\ and\ \citenamefont {H\"ofling}}]{He2017}%
	\BibitemOpen
	\bibfield  {author} {\bibinfo {author} {\bibfnamefont {Y.}~\bibnamefont
			{He}}, \bibinfo {author} {\bibfnamefont {Y.-M.}\ \bibnamefont {He}}, \bibinfo
		{author} {\bibfnamefont {Y.-J.}\ \bibnamefont {Wei}}, \bibinfo {author}
		{\bibfnamefont {X.}~\bibnamefont {Jiang}}, \bibinfo {author} {\bibfnamefont
			{K.}~\bibnamefont {Chen}}, \bibinfo {author} {\bibfnamefont {C.-Y.}\
			\bibnamefont {Lu}}, \bibinfo {author} {\bibfnamefont {J.-W.}\ \bibnamefont
			{Pan}}, \bibinfo {author} {\bibfnamefont {C.}~\bibnamefont {Schneider}},
		\bibinfo {author} {\bibfnamefont {M.}~\bibnamefont {Kamp}}, \ and\ \bibinfo
		{author} {\bibfnamefont {S.}~\bibnamefont {H\"ofling}},\ }\href {\doibase
		10.1103/PhysRevLett.119.060501} {\bibfield  {journal} {\bibinfo  {journal}
			{Phys. Rev. Lett.}\ }\textbf {\bibinfo {volume} {119}},\ \bibinfo {pages}
		{060501} (\bibinfo {year} {2017})}\BibitemShut {NoStop}%
	\bibitem [{\citenamefont {Hanson}\ \emph {et~al.}(2007)\citenamefont {Hanson},
		\citenamefont {Kouwenhoven}, \citenamefont {Petta}, \citenamefont {Tarucha},\
		and\ \citenamefont {Vandersypen}}]{Hanson2007}%
	\BibitemOpen
	\bibfield  {author} {\bibinfo {author} {\bibfnamefont {R.}~\bibnamefont
			{Hanson}}, \bibinfo {author} {\bibfnamefont {L.~P.}\ \bibnamefont
			{Kouwenhoven}}, \bibinfo {author} {\bibfnamefont {J.~R.}\ \bibnamefont
			{Petta}}, \bibinfo {author} {\bibfnamefont {S.}~\bibnamefont {Tarucha}}, \
		and\ \bibinfo {author} {\bibfnamefont {L.~M.~K.}\ \bibnamefont
			{Vandersypen}},\ }\href {\doibase 10.1103/RevModPhys.79.1217} {\bibfield
		{journal} {\bibinfo  {journal} {Rev. Mod. Phys.}\ }\textbf {\bibinfo {volume}
			{79}},\ \bibinfo {pages} {1217} (\bibinfo {year} {2007})}\BibitemShut
	{NoStop}%
	\bibitem [{\citenamefont {Bluhm}\ \emph {et~al.}(2011)\citenamefont {Bluhm},
		\citenamefont {Foletti}, \citenamefont {Neder}, \citenamefont {Rudner},
		\citenamefont {Mahalu}, \citenamefont {Umansky},\ and\ \citenamefont
		{Yacoby}}]{Bluhm2011}%
	\BibitemOpen
	\bibfield  {author} {\bibinfo {author} {\bibfnamefont {H.}~\bibnamefont
			{Bluhm}}, \bibinfo {author} {\bibfnamefont {S.}~\bibnamefont {Foletti}},
		\bibinfo {author} {\bibfnamefont {I.}~\bibnamefont {Neder}}, \bibinfo
		{author} {\bibfnamefont {M.}~\bibnamefont {Rudner}}, \bibinfo {author}
		{\bibfnamefont {D.}~\bibnamefont {Mahalu}}, \bibinfo {author} {\bibfnamefont
			{V.}~\bibnamefont {Umansky}}, \ and\ \bibinfo {author} {\bibfnamefont
			{A.}~\bibnamefont {Yacoby}},\ }\href {\doibase 10.1038/nphys1856} {\bibfield
		{journal} {\bibinfo  {journal} {Nature Physics}\ }\textbf {\bibinfo {volume}
			{7}},\ \bibinfo {pages} {109} (\bibinfo {year} {2011})}\BibitemShut {NoStop}%
	\bibitem [{\citenamefont {Granger}\ \emph {et~al.}(2012)\citenamefont
		{Granger}, \citenamefont {Taubert}, \citenamefont {Young}, \citenamefont
		{Gaudreau}, \citenamefont {Kam}, \citenamefont {Studenikin}, \citenamefont
		{Zawadzki}, \citenamefont {Harbusch}, \citenamefont {Schuh}, \citenamefont
		{Wegscheider}, \citenamefont {Wasilewski}, \citenamefont {Clerk},
		\citenamefont {Ludwig},\ and\ \citenamefont {Sachrajda}}]{Granger2012}%
	\BibitemOpen
	\bibfield  {author} {\bibinfo {author} {\bibfnamefont {G.}~\bibnamefont
			{Granger}}, \bibinfo {author} {\bibfnamefont {D.}~\bibnamefont {Taubert}},
		\bibinfo {author} {\bibfnamefont {C.~E.}\ \bibnamefont {Young}}, \bibinfo
		{author} {\bibfnamefont {L.}~\bibnamefont {Gaudreau}}, \bibinfo {author}
		{\bibfnamefont {A.}~\bibnamefont {Kam}}, \bibinfo {author} {\bibfnamefont
			{S.~A.}\ \bibnamefont {Studenikin}}, \bibinfo {author} {\bibfnamefont
			{P.}~\bibnamefont {Zawadzki}}, \bibinfo {author} {\bibfnamefont
			{D.}~\bibnamefont {Harbusch}}, \bibinfo {author} {\bibfnamefont
			{D.}~\bibnamefont {Schuh}}, \bibinfo {author} {\bibfnamefont
			{W.}~\bibnamefont {Wegscheider}}, \bibinfo {author} {\bibfnamefont {Z.~R.}\
			\bibnamefont {Wasilewski}}, \bibinfo {author} {\bibfnamefont {A.~A.}\
			\bibnamefont {Clerk}}, \bibinfo {author} {\bibfnamefont {S.}~\bibnamefont
			{Ludwig}}, \ and\ \bibinfo {author} {\bibfnamefont {A.~S.}\ \bibnamefont
			{Sachrajda}},\ }\href {\doibase 10.1038/nphys2326} {\bibfield  {journal}
		{\bibinfo  {journal} {Nature Physics}\ }\textbf {\bibinfo {volume} {8}},\
		\bibinfo {pages} {522} (\bibinfo {year} {2012})}\BibitemShut {NoStop}%
	\bibitem [{\citenamefont {S\'anchez}\ and\ \citenamefont
		{Platero}(2013)}]{Sanchez2013}%
	\BibitemOpen
	\bibfield  {author} {\bibinfo {author} {\bibfnamefont {R.}~\bibnamefont
			{S\'anchez}}\ and\ \bibinfo {author} {\bibfnamefont {G.}~\bibnamefont
			{Platero}},\ }\href {\doibase 10.1103/PhysRevB.87.081305} {\bibfield
		{journal} {\bibinfo  {journal} {Phys. Rev. B}\ }\textbf {\bibinfo {volume}
			{87}},\ \bibinfo {pages} {081305} (\bibinfo {year} {2013})}\BibitemShut
	{NoStop}%
	\bibitem [{\citenamefont {Ito}\ \emph {et~al.}(2016)\citenamefont {Ito},
		\citenamefont {Otsuka}, \citenamefont {Amaha}, \citenamefont {Delbecq},
		\citenamefont {Nakajima}, \citenamefont {Yoneda}, \citenamefont {Takeda},
		\citenamefont {Allison}, \citenamefont {Noiri}, \citenamefont {Kawasaki},\
		and\ \citenamefont {Tarucha}}]{Ito2016}%
	\BibitemOpen
	\bibfield  {author} {\bibinfo {author} {\bibfnamefont {T.}~\bibnamefont
			{Ito}}, \bibinfo {author} {\bibfnamefont {T.}~\bibnamefont {Otsuka}},
		\bibinfo {author} {\bibfnamefont {S.}~\bibnamefont {Amaha}}, \bibinfo
		{author} {\bibfnamefont {M.~R.}\ \bibnamefont {Delbecq}}, \bibinfo {author}
		{\bibfnamefont {T.}~\bibnamefont {Nakajima}}, \bibinfo {author}
		{\bibfnamefont {J.}~\bibnamefont {Yoneda}}, \bibinfo {author} {\bibfnamefont
			{K.}~\bibnamefont {Takeda}}, \bibinfo {author} {\bibfnamefont
			{G.}~\bibnamefont {Allison}}, \bibinfo {author} {\bibfnamefont
			{A.}~\bibnamefont {Noiri}}, \bibinfo {author} {\bibfnamefont
			{K.}~\bibnamefont {Kawasaki}}, \ and\ \bibinfo {author} {\bibfnamefont
			{S.}~\bibnamefont {Tarucha}},\ }\href {\doibase 10.1038/srep39113} {\bibfield
		{journal} {\bibinfo  {journal} {Scientific Reports}\ }\textbf {\bibinfo
			{volume} {6}} (\bibinfo {year} {2016}),\ 10.1038/srep39113}\BibitemShut
	{NoStop}%
	\bibitem [{\citenamefont {Zajac}\ \emph {et~al.}(2016)\citenamefont {Zajac},
		\citenamefont {Hazard}, \citenamefont {Mi}, \citenamefont {Nielsen},\ and\
		\citenamefont {Petta}}]{Zajac2016}%
	\BibitemOpen
	\bibfield  {author} {\bibinfo {author} {\bibfnamefont {D.~M.}\ \bibnamefont
			{Zajac}}, \bibinfo {author} {\bibfnamefont {T.~M.}\ \bibnamefont {Hazard}},
		\bibinfo {author} {\bibfnamefont {X.}~\bibnamefont {Mi}}, \bibinfo {author}
		{\bibfnamefont {E.}~\bibnamefont {Nielsen}}, \ and\ \bibinfo {author}
		{\bibfnamefont {J.~R.}\ \bibnamefont {Petta}},\ }\href {\doibase
		10.1103/PhysRevApplied.6.054013} {\bibfield  {journal} {\bibinfo  {journal}
			{Phys. Rev. Applied}\ }\textbf {\bibinfo {volume} {6}},\ \bibinfo {pages}
		{054013} (\bibinfo {year} {2016})}\BibitemShut {NoStop}%
	\bibitem [{\citenamefont {Fujita}\ \emph {et~al.}(2017)\citenamefont {Fujita},
		\citenamefont {Baart}, \citenamefont {Reichl}, \citenamefont {Wegscheider},\
		and\ \citenamefont {Vandersypen}}]{Fujita2017}%
	\BibitemOpen
	\bibfield  {author} {\bibinfo {author} {\bibfnamefont {T.}~\bibnamefont
			{Fujita}}, \bibinfo {author} {\bibfnamefont {T.~A.}\ \bibnamefont {Baart}},
		\bibinfo {author} {\bibfnamefont {C.}~\bibnamefont {Reichl}}, \bibinfo
		{author} {\bibfnamefont {W.}~\bibnamefont {Wegscheider}}, \ and\ \bibinfo
		{author} {\bibfnamefont {L.~M.~K.}\ \bibnamefont {Vandersypen}},\ }\href
	{\doibase 10.1038/s41534-017-0024-4} {\bibfield  {journal} {\bibinfo
			{journal} {npj Quantum Information}\ }\textbf {\bibinfo {volume} {3}}
		(\bibinfo {year} {2017}),\ 10.1038/s41534-017-0024-4}\BibitemShut {NoStop}%
	\bibitem [{\citenamefont {Korkusinski}\ \emph {et~al.}(2007)\citenamefont
		{Korkusinski}, \citenamefont {Gimenez}, \citenamefont {Hawrylak},
		\citenamefont {Gaudreau}, \citenamefont {Studenikin},\ and\ \citenamefont
		{Sachrajda}}]{Korkusinski2007}%
	\BibitemOpen
	\bibfield  {author} {\bibinfo {author} {\bibfnamefont {M.}~\bibnamefont
			{Korkusinski}}, \bibinfo {author} {\bibfnamefont {I.~P.}\ \bibnamefont
			{Gimenez}}, \bibinfo {author} {\bibfnamefont {P.}~\bibnamefont {Hawrylak}},
		\bibinfo {author} {\bibfnamefont {L.}~\bibnamefont {Gaudreau}}, \bibinfo
		{author} {\bibfnamefont {S.~A.}\ \bibnamefont {Studenikin}}, \ and\ \bibinfo
		{author} {\bibfnamefont {A.~S.}\ \bibnamefont {Sachrajda}},\ }\href {\doibase
		10.1103/PhysRevB.75.115301} {\bibfield  {journal} {\bibinfo  {journal} {Phys.
				Rev. B}\ }\textbf {\bibinfo {volume} {75}},\ \bibinfo {pages} {115301}
		(\bibinfo {year} {2007})}\BibitemShut {NoStop}%
	\bibitem [{\citenamefont {Kotzian}\ \emph {et~al.}(2016)\citenamefont
		{Kotzian}, \citenamefont {Gallego-Marcos}, \citenamefont {Platero},\ and\
		\citenamefont {Haug}}]{Kotzian2016}%
	\BibitemOpen
	\bibfield  {author} {\bibinfo {author} {\bibfnamefont {M.}~\bibnamefont
			{Kotzian}}, \bibinfo {author} {\bibfnamefont {F.}~\bibnamefont
			{Gallego-Marcos}}, \bibinfo {author} {\bibfnamefont {G.}~\bibnamefont
			{Platero}}, \ and\ \bibinfo {author} {\bibfnamefont {R.~J.}\ \bibnamefont
			{Haug}},\ }\href {\doibase 10.1103/PhysRevB.94.035442} {\bibfield  {journal}
		{\bibinfo  {journal} {Phys. Rev. B}\ }\textbf {\bibinfo {volume} {94}},\
		\bibinfo {pages} {035442} (\bibinfo {year} {2016})}\BibitemShut {NoStop}%
	\bibitem [{\citenamefont {Gaudreau}\ \emph {et~al.}(2011)\citenamefont
		{Gaudreau}, \citenamefont {Granger}, \citenamefont {Kam}, \citenamefont
		{Aers}, \citenamefont {Studenikin}, \citenamefont {Zawadzki}, \citenamefont
		{Pioro-Ladri{\`{e}}re}, \citenamefont {Wasilewski},\ and\ \citenamefont
		{Sachrajda}}]{Gaudreau2011}%
	\BibitemOpen
	\bibfield  {author} {\bibinfo {author} {\bibfnamefont {L.}~\bibnamefont
			{Gaudreau}}, \bibinfo {author} {\bibfnamefont {G.}~\bibnamefont {Granger}},
		\bibinfo {author} {\bibfnamefont {A.}~\bibnamefont {Kam}}, \bibinfo {author}
		{\bibfnamefont {G.~C.}\ \bibnamefont {Aers}}, \bibinfo {author}
		{\bibfnamefont {S.~A.}\ \bibnamefont {Studenikin}}, \bibinfo {author}
		{\bibfnamefont {P.}~\bibnamefont {Zawadzki}}, \bibinfo {author}
		{\bibfnamefont {M.}~\bibnamefont {Pioro-Ladri{\`{e}}re}}, \bibinfo {author}
		{\bibfnamefont {Z.~R.}\ \bibnamefont {Wasilewski}}, \ and\ \bibinfo {author}
		{\bibfnamefont {A.~S.}\ \bibnamefont {Sachrajda}},\ }\href {\doibase
		10.1038/nphys2149} {\bibfield  {journal} {\bibinfo  {journal} {Nature
				Physics}\ }\textbf {\bibinfo {volume} {8}},\ \bibinfo {pages} {54} (\bibinfo
		{year} {2011})}\BibitemShut {NoStop}%
	\bibitem [{\citenamefont {Medford}\ \emph {et~al.}(2013)\citenamefont
		{Medford}, \citenamefont {Beil}, \citenamefont {Taylor}, \citenamefont
		{Bartlett}, \citenamefont {Doherty}, \citenamefont {Rashba}, \citenamefont
		{DiVincenzo}, \citenamefont {Lu}, \citenamefont {Gossard},\ and\
		\citenamefont {Marcus}}]{Medford2013}%
	\BibitemOpen
	\bibfield  {author} {\bibinfo {author} {\bibfnamefont {J.}~\bibnamefont
			{Medford}}, \bibinfo {author} {\bibfnamefont {J.}~\bibnamefont {Beil}},
		\bibinfo {author} {\bibfnamefont {J.~M.}\ \bibnamefont {Taylor}}, \bibinfo
		{author} {\bibfnamefont {S.~D.}\ \bibnamefont {Bartlett}}, \bibinfo {author}
		{\bibfnamefont {A.~C.}\ \bibnamefont {Doherty}}, \bibinfo {author}
		{\bibfnamefont {E.~I.}\ \bibnamefont {Rashba}}, \bibinfo {author}
		{\bibfnamefont {D.~P.}\ \bibnamefont {DiVincenzo}}, \bibinfo {author}
		{\bibfnamefont {H.}~\bibnamefont {Lu}}, \bibinfo {author} {\bibfnamefont
			{A.~C.}\ \bibnamefont {Gossard}}, \ and\ \bibinfo {author} {\bibfnamefont
			{C.~M.}\ \bibnamefont {Marcus}},\ }\href {\doibase 10.1038/nnano.2013.168}
	{\bibfield  {journal} {\bibinfo  {journal} {Nature Nanotechnology}\ }\textbf
		{\bibinfo {volume} {8}},\ \bibinfo {pages} {654} (\bibinfo {year}
		{2013})}\BibitemShut {NoStop}%
	\bibitem [{\citenamefont {Russ}\ and\ \citenamefont
		{Burkard}(2017)}]{Russ2017}%
	\BibitemOpen
	\bibfield  {author} {\bibinfo {author} {\bibfnamefont {M.}~\bibnamefont
			{Russ}}\ and\ \bibinfo {author} {\bibfnamefont {G.}~\bibnamefont {Burkard}},\
	}\href {\doibase 10.1088/1361-648x/aa761f} {\bibfield  {journal} {\bibinfo
			{journal} {Journal of Physics: Condensed Matter}\ }\textbf {\bibinfo {volume}
			{29}},\ \bibinfo {pages} {393001} (\bibinfo {year} {2017})}\BibitemShut
	{NoStop}%
	\bibitem [{\citenamefont {Busl}\ \emph {et~al.}(2013)\citenamefont {Busl},
		\citenamefont {Granger}, \citenamefont {Gaudreau}, \citenamefont
		{S{\'{a}}nchez}, \citenamefont {Kam}, \citenamefont {Pioro-Ladri{\`{e}}re},
		\citenamefont {Studenikin}, \citenamefont {Zawadzki}, \citenamefont
		{Wasilewski}, \citenamefont {Sachrajda},\ and\ \citenamefont
		{Platero}}]{Busl2013}%
	\BibitemOpen
	\bibfield  {author} {\bibinfo {author} {\bibfnamefont {M.}~\bibnamefont
			{Busl}}, \bibinfo {author} {\bibfnamefont {G.}~\bibnamefont {Granger}},
		\bibinfo {author} {\bibfnamefont {L.}~\bibnamefont {Gaudreau}}, \bibinfo
		{author} {\bibfnamefont {R.}~\bibnamefont {S{\'{a}}nchez}}, \bibinfo {author}
		{\bibfnamefont {A.}~\bibnamefont {Kam}}, \bibinfo {author} {\bibfnamefont
			{M.}~\bibnamefont {Pioro-Ladri{\`{e}}re}}, \bibinfo {author} {\bibfnamefont
			{S.~A.}\ \bibnamefont {Studenikin}}, \bibinfo {author} {\bibfnamefont
			{P.}~\bibnamefont {Zawadzki}}, \bibinfo {author} {\bibfnamefont {Z.~R.}\
			\bibnamefont {Wasilewski}}, \bibinfo {author} {\bibfnamefont {A.~S.}\
			\bibnamefont {Sachrajda}}, \ and\ \bibinfo {author} {\bibfnamefont
			{G.}~\bibnamefont {Platero}},\ }\href {\doibase 10.1038/nnano.2013.7}
	{\bibfield  {journal} {\bibinfo  {journal} {Nature Nanotechnology}\ }\textbf
		{\bibinfo {volume} {8}},\ \bibinfo {pages} {261} (\bibinfo {year}
		{2013})}\BibitemShut {NoStop}%
	\bibitem [{\citenamefont {Braakman}\ \emph {et~al.}(2013)\citenamefont
		{Braakman}, \citenamefont {Barthelemy}, \citenamefont {Reichl}, \citenamefont
		{Wegscheider},\ and\ \citenamefont {Vandersypen}}]{Braakman2013}%
	\BibitemOpen
	\bibfield  {author} {\bibinfo {author} {\bibfnamefont {F.~R.}\ \bibnamefont
			{Braakman}}, \bibinfo {author} {\bibfnamefont {P.}~\bibnamefont
			{Barthelemy}}, \bibinfo {author} {\bibfnamefont {C.}~\bibnamefont {Reichl}},
		\bibinfo {author} {\bibfnamefont {W.}~\bibnamefont {Wegscheider}}, \ and\
		\bibinfo {author} {\bibfnamefont {L.~M.~K.}\ \bibnamefont {Vandersypen}},\
	}\href {\doibase 10.1038/nnano.2013.67} {\bibfield  {journal} {\bibinfo
			{journal} {Nature Nanotechnology}\ }\textbf {\bibinfo {volume} {8}},\
		\bibinfo {pages} {432} (\bibinfo {year} {2013})}\BibitemShut {NoStop}%
	\bibitem [{\citenamefont {S\'anchez}\ \emph {et~al.}(2014)\citenamefont
		{S\'anchez}, \citenamefont {Granger}, \citenamefont {Gaudreau}, \citenamefont
		{Kam}, \citenamefont {Pioro-Ladri\`ere}, \citenamefont {Studenikin},
		\citenamefont {Zawadzki}, \citenamefont {Sachrajda},\ and\ \citenamefont
		{Platero}}]{Sanchez2014}%
	\BibitemOpen
	\bibfield  {author} {\bibinfo {author} {\bibfnamefont {R.}~\bibnamefont
			{S\'anchez}}, \bibinfo {author} {\bibfnamefont {G.}~\bibnamefont {Granger}},
		\bibinfo {author} {\bibfnamefont {L.}~\bibnamefont {Gaudreau}}, \bibinfo
		{author} {\bibfnamefont {A.}~\bibnamefont {Kam}}, \bibinfo {author}
		{\bibfnamefont {M.}~\bibnamefont {Pioro-Ladri\`ere}}, \bibinfo {author}
		{\bibfnamefont {S.~A.}\ \bibnamefont {Studenikin}}, \bibinfo {author}
		{\bibfnamefont {P.}~\bibnamefont {Zawadzki}}, \bibinfo {author}
		{\bibfnamefont {A.~S.}\ \bibnamefont {Sachrajda}}, \ and\ \bibinfo {author}
		{\bibfnamefont {G.}~\bibnamefont {Platero}},\ }\href {\doibase
		10.1103/PhysRevLett.112.176803} {\bibfield  {journal} {\bibinfo  {journal}
			{Phys. Rev. Lett.}\ }\textbf {\bibinfo {volume} {112}},\ \bibinfo {pages}
		{176803} (\bibinfo {year} {2014})}\BibitemShut {NoStop}%
	\bibitem [{\citenamefont {Wang}\ \emph {et~al.}(2017)\citenamefont {Wang},
		\citenamefont {Huang}, \citenamefont {Huang}, \citenamefont {Pan},
		\citenamefont {Zhao},\ and\ \citenamefont {Xu}}]{Wang2017}%
	\BibitemOpen
	\bibfield  {author} {\bibinfo {author} {\bibfnamefont {J.-Y.}\ \bibnamefont
			{Wang}}, \bibinfo {author} {\bibfnamefont {S.}~\bibnamefont {Huang}},
		\bibinfo {author} {\bibfnamefont {G.-Y.}\ \bibnamefont {Huang}}, \bibinfo
		{author} {\bibfnamefont {D.}~\bibnamefont {Pan}}, \bibinfo {author}
		{\bibfnamefont {J.}~\bibnamefont {Zhao}}, \ and\ \bibinfo {author}
		{\bibfnamefont {H.~Q.}\ \bibnamefont {Xu}},\ }\href {\doibase
		10.1021/acs.nanolett.7b00927} {\bibfield  {journal} {\bibinfo  {journal}
			{Nano Letters}\ }\textbf {\bibinfo {volume} {17}},\ \bibinfo {pages} {4158}
		(\bibinfo {year} {2017})}\BibitemShut {NoStop}%
	\bibitem [{\citenamefont {Ban}\ \emph {et~al.}(2018)\citenamefont {Ban},
		\citenamefont {Chen},\ and\ \citenamefont {Platero}}]{Ban2018}%
	\BibitemOpen
	\bibfield  {author} {\bibinfo {author} {\bibfnamefont {Y.}~\bibnamefont
			{Ban}}, \bibinfo {author} {\bibfnamefont {X.}~\bibnamefont {Chen}}, \ and\
		\bibinfo {author} {\bibfnamefont {G.}~\bibnamefont {Platero}},\ }\href
	{\doibase 10.1088/1361-6528/aae0ce} {\bibfield  {journal} {\bibinfo
			{journal} {Nanotechnology}\ }\textbf {\bibinfo {volume} {29}},\ \bibinfo
		{pages} {505201} (\bibinfo {year} {2018})}\BibitemShut {NoStop}%
	\bibitem [{\citenamefont {Rahman}\ \emph {et~al.}(2010)\citenamefont {Rahman},
		\citenamefont {Muller}, \citenamefont {Levy}, \citenamefont {Carroll},
		\citenamefont {Klimeck}, \citenamefont {Greentree},\ and\ \citenamefont
		{Hollenberg}}]{Rahman2010}%
	\BibitemOpen
	\bibfield  {author} {\bibinfo {author} {\bibfnamefont {R.}~\bibnamefont
			{Rahman}}, \bibinfo {author} {\bibfnamefont {R.~P.}\ \bibnamefont {Muller}},
		\bibinfo {author} {\bibfnamefont {J.~E.}\ \bibnamefont {Levy}}, \bibinfo
		{author} {\bibfnamefont {M.~S.}\ \bibnamefont {Carroll}}, \bibinfo {author}
		{\bibfnamefont {G.}~\bibnamefont {Klimeck}}, \bibinfo {author} {\bibfnamefont
			{A.~D.}\ \bibnamefont {Greentree}}, \ and\ \bibinfo {author} {\bibfnamefont
			{L.~C.~L.}\ \bibnamefont {Hollenberg}},\ }\href {\doibase
		10.1103/PhysRevB.82.155315} {\bibfield  {journal} {\bibinfo  {journal} {Phys.
				Rev. B}\ }\textbf {\bibinfo {volume} {82}},\ \bibinfo {pages} {155315}
		(\bibinfo {year} {2010})}\BibitemShut {NoStop}%
	\bibitem [{\citenamefont {Schreiber}\ \emph {et~al.}(2011)\citenamefont
		{Schreiber}, \citenamefont {Braakman}, \citenamefont {Meunier}, \citenamefont
		{Calado}, \citenamefont {Danon}, \citenamefont {Taylor}, \citenamefont
		{Wegscheider},\ and\ \citenamefont {Vandersypen}}]{Schreiber2011}%
	\BibitemOpen
	\bibfield  {author} {\bibinfo {author} {\bibfnamefont {L.}~\bibnamefont
			{Schreiber}}, \bibinfo {author} {\bibfnamefont {F.}~\bibnamefont {Braakman}},
		\bibinfo {author} {\bibfnamefont {T.}~\bibnamefont {Meunier}}, \bibinfo
		{author} {\bibfnamefont {V.}~\bibnamefont {Calado}}, \bibinfo {author}
		{\bibfnamefont {J.}~\bibnamefont {Danon}}, \bibinfo {author} {\bibfnamefont
			{J.}~\bibnamefont {Taylor}}, \bibinfo {author} {\bibfnamefont
			{W.}~\bibnamefont {Wegscheider}}, \ and\ \bibinfo {author} {\bibfnamefont
			{L.}~\bibnamefont {Vandersypen}},\ }\href {\doibase 10.1038/ncomms1561}
	{\bibfield  {journal} {\bibinfo  {journal} {Nature Communications}\ }\textbf
		{\bibinfo {volume} {2}},\ \bibinfo {pages} {556} (\bibinfo {year}
		{2011})}\BibitemShut {NoStop}%
	\bibitem [{\citenamefont {Gallego-Marcos}\ \emph {et~al.}(2015)\citenamefont
		{Gallego-Marcos}, \citenamefont {S{\'{a}}nchez},\ and\ \citenamefont
		{Platero}}]{GallegoMarcos2015}%
	\BibitemOpen
	\bibfield  {author} {\bibinfo {author} {\bibfnamefont {F.}~\bibnamefont
			{Gallego-Marcos}}, \bibinfo {author} {\bibfnamefont {R.}~\bibnamefont
			{S{\'{a}}nchez}}, \ and\ \bibinfo {author} {\bibfnamefont {G.}~\bibnamefont
			{Platero}},\ }\href {\doibase 10.1063/1.4913834} {\bibfield  {journal}
		{\bibinfo  {journal} {Journal of Applied Physics}\ }\textbf {\bibinfo
			{volume} {117}},\ \bibinfo {pages} {112808} (\bibinfo {year}
		{2015})}\BibitemShut {NoStop}%
	\bibitem [{\citenamefont {Gallego-Marcos}\ and\ \citenamefont
		{Platero}(2017)}]{GallegoMarcos2017}%
	\BibitemOpen
	\bibfield  {author} {\bibinfo {author} {\bibfnamefont {F.}~\bibnamefont
			{Gallego-Marcos}}\ and\ \bibinfo {author} {\bibfnamefont {G.}~\bibnamefont
			{Platero}},\ }\href {\doibase 10.1103/PhysRevB.95.075301} {\bibfield
		{journal} {\bibinfo  {journal} {Phys. Rev. B}\ }\textbf {\bibinfo {volume}
			{95}},\ \bibinfo {pages} {075301} (\bibinfo {year} {2017})}\BibitemShut
	{NoStop}%
	\bibitem [{\citenamefont {Gallego-Marcos}\ \emph {et~al.}(2016)\citenamefont
		{Gallego-Marcos}, \citenamefont {S\'anchez},\ and\ \citenamefont
		{Platero}}]{GallegoMarcos2016}%
	\BibitemOpen
	\bibfield  {author} {\bibinfo {author} {\bibfnamefont {F.}~\bibnamefont
			{Gallego-Marcos}}, \bibinfo {author} {\bibfnamefont {R.}~\bibnamefont
			{S\'anchez}}, \ and\ \bibinfo {author} {\bibfnamefont {G.}~\bibnamefont
			{Platero}},\ }\href {\doibase 10.1103/PhysRevB.93.075424} {\bibfield
		{journal} {\bibinfo  {journal} {Phys. Rev. B}\ }\textbf {\bibinfo {volume}
			{93}},\ \bibinfo {pages} {075424} (\bibinfo {year} {2016})}\BibitemShut
	{NoStop}%
	\bibitem [{\citenamefont {Wardrop}\ and\ \citenamefont
		{Doherty}(2014)}]{Wardrop2014}%
	\BibitemOpen
	\bibfield  {author} {\bibinfo {author} {\bibfnamefont {M.~P.}\ \bibnamefont
			{Wardrop}}\ and\ \bibinfo {author} {\bibfnamefont {A.~C.}\ \bibnamefont
			{Doherty}},\ }\href {\doibase 10.1103/PhysRevB.90.045418} {\bibfield
		{journal} {\bibinfo  {journal} {Phys. Rev. B}\ }\textbf {\bibinfo {volume}
			{90}},\ \bibinfo {pages} {045418} (\bibinfo {year} {2014})}\BibitemShut
	{NoStop}%
	\bibitem [{\citenamefont {Feng}\ \emph {et~al.}(2018)\citenamefont {Feng},
		\citenamefont {Kwong}, \citenamefont {Koh},\ and\ \citenamefont
		{Kwek}}]{Feng2018}%
	\BibitemOpen
	\bibfield  {author} {\bibinfo {author} {\bibfnamefont {M.}~\bibnamefont
			{Feng}}, \bibinfo {author} {\bibfnamefont {C.~J.}\ \bibnamefont {Kwong}},
		\bibinfo {author} {\bibfnamefont {T.~S.}\ \bibnamefont {Koh}}, \ and\
		\bibinfo {author} {\bibfnamefont {L.~C.}\ \bibnamefont {Kwek}},\ }\href
	{\doibase 10.1103/PhysRevB.97.245428} {\bibfield  {journal} {\bibinfo
			{journal} {Phys. Rev. B}\ }\textbf {\bibinfo {volume} {97}},\ \bibinfo
		{pages} {245428} (\bibinfo {year} {2018})}\BibitemShut {NoStop}%
	\bibitem [{\citenamefont {Goldin}\ and\ \citenamefont
		{Avishai}(2000)}]{Goldin2000}%
	\BibitemOpen
	\bibfield  {author} {\bibinfo {author} {\bibfnamefont {Y.}~\bibnamefont
			{Goldin}}\ and\ \bibinfo {author} {\bibfnamefont {Y.}~\bibnamefont
			{Avishai}},\ }\href {\doibase 10.1103/PhysRevB.61.16750} {\bibfield
		{journal} {\bibinfo  {journal} {Phys. Rev. B}\ }\textbf {\bibinfo {volume}
			{61}},\ \bibinfo {pages} {16750} (\bibinfo {year} {2000})}\BibitemShut
	{NoStop}%
	\bibitem [{\citenamefont {Fei}\ \emph {et~al.}(2015)\citenamefont {Fei},
		\citenamefont {Hung}, \citenamefont {Koh}, \citenamefont {Shim},
		\citenamefont {Coppersmith}, \citenamefont {Hu},\ and\ \citenamefont
		{Friesen}}]{Fei2015}%
	\BibitemOpen
	\bibfield  {author} {\bibinfo {author} {\bibfnamefont {J.}~\bibnamefont
			{Fei}}, \bibinfo {author} {\bibfnamefont {J.-T.}\ \bibnamefont {Hung}},
		\bibinfo {author} {\bibfnamefont {T.~S.}\ \bibnamefont {Koh}}, \bibinfo
		{author} {\bibfnamefont {Y.-P.}\ \bibnamefont {Shim}}, \bibinfo {author}
		{\bibfnamefont {S.~N.}\ \bibnamefont {Coppersmith}}, \bibinfo {author}
		{\bibfnamefont {X.}~\bibnamefont {Hu}}, \ and\ \bibinfo {author}
		{\bibfnamefont {M.}~\bibnamefont {Friesen}},\ }\href {\doibase
		10.1103/PhysRevB.91.205434} {\bibfield  {journal} {\bibinfo  {journal} {Phys.
				Rev. B}\ }\textbf {\bibinfo {volume} {91}},\ \bibinfo {pages} {205434}
		(\bibinfo {year} {2015})}\BibitemShut {NoStop}%
	\bibitem [{\citenamefont {Martins}\ \emph {et~al.}(2016)\citenamefont
		{Martins}, \citenamefont {Malinowski}, \citenamefont {Nissen}, \citenamefont
		{Barnes}, \citenamefont {Fallahi}, \citenamefont {Gardner}, \citenamefont
		{Manfra}, \citenamefont {Marcus},\ and\ \citenamefont
		{Kuemmeth}}]{Martins2016}%
	\BibitemOpen
	\bibfield  {author} {\bibinfo {author} {\bibfnamefont {F.}~\bibnamefont
			{Martins}}, \bibinfo {author} {\bibfnamefont {F.~K.}\ \bibnamefont
			{Malinowski}}, \bibinfo {author} {\bibfnamefont {P.~D.}\ \bibnamefont
			{Nissen}}, \bibinfo {author} {\bibfnamefont {E.}~\bibnamefont {Barnes}},
		\bibinfo {author} {\bibfnamefont {S.}~\bibnamefont {Fallahi}}, \bibinfo
		{author} {\bibfnamefont {G.~C.}\ \bibnamefont {Gardner}}, \bibinfo {author}
		{\bibfnamefont {M.~J.}\ \bibnamefont {Manfra}}, \bibinfo {author}
		{\bibfnamefont {C.~M.}\ \bibnamefont {Marcus}}, \ and\ \bibinfo {author}
		{\bibfnamefont {F.}~\bibnamefont {Kuemmeth}},\ }\href {\doibase
		10.1103/PhysRevLett.116.116801} {\bibfield  {journal} {\bibinfo  {journal}
			{Phys. Rev. Lett.}\ }\textbf {\bibinfo {volume} {116}},\ \bibinfo {pages}
		{116801} (\bibinfo {year} {2016})}\BibitemShut {NoStop}%
	\bibitem [{Note1()}]{Note1}%
	\BibitemOpen
	\bibinfo {note} {See Supplementary information.}\BibitemShut {Stop}%
	\bibitem [{\citenamefont {Petersen}\ \emph {et~al.}(2013)\citenamefont
		{Petersen}, \citenamefont {Hoffmann}, \citenamefont {Schuh}, \citenamefont
		{Wegscheider}, \citenamefont {Giedke},\ and\ \citenamefont
		{Ludwig}}]{Petersen2013}%
	\BibitemOpen
	\bibfield  {author} {\bibinfo {author} {\bibfnamefont {G.}~\bibnamefont
			{Petersen}}, \bibinfo {author} {\bibfnamefont {E.~A.}\ \bibnamefont
			{Hoffmann}}, \bibinfo {author} {\bibfnamefont {D.}~\bibnamefont {Schuh}},
		\bibinfo {author} {\bibfnamefont {W.}~\bibnamefont {Wegscheider}}, \bibinfo
		{author} {\bibfnamefont {G.}~\bibnamefont {Giedke}}, \ and\ \bibinfo {author}
		{\bibfnamefont {S.}~\bibnamefont {Ludwig}},\ }\href {\doibase
		10.1103/PhysRevLett.110.177602} {\bibfield  {journal} {\bibinfo  {journal}
			{Phys. Rev. Lett.}\ }\textbf {\bibinfo {volume} {110}},\ \bibinfo {pages}
		{177602} (\bibinfo {year} {2013})}\BibitemShut {NoStop}%
	\bibitem [{\citenamefont {Forster}\ \emph {et~al.}(2015)\citenamefont
		{Forster}, \citenamefont {M\"uhlbacher}, \citenamefont {Schuh}, \citenamefont
		{Wegscheider},\ and\ \citenamefont {Ludwig}}]{Forster2015}%
	\BibitemOpen
	\bibfield  {author} {\bibinfo {author} {\bibfnamefont {F.}~\bibnamefont
			{Forster}}, \bibinfo {author} {\bibfnamefont {M.}~\bibnamefont
			{M\"uhlbacher}}, \bibinfo {author} {\bibfnamefont {D.}~\bibnamefont {Schuh}},
		\bibinfo {author} {\bibfnamefont {W.}~\bibnamefont {Wegscheider}}, \ and\
		\bibinfo {author} {\bibfnamefont {S.}~\bibnamefont {Ludwig}},\ }\href
	{\doibase 10.1103/PhysRevB.91.195417} {\bibfield  {journal} {\bibinfo
			{journal} {Phys. Rev. B}\ }\textbf {\bibinfo {volume} {91}},\ \bibinfo
		{pages} {195417} (\bibinfo {year} {2015})}\BibitemShut {NoStop}%
	\bibitem [{\citenamefont {Yoneda}\ \emph {et~al.}(2015)\citenamefont {Yoneda},
		\citenamefont {Otsuka}, \citenamefont {Takakura}, \citenamefont
		{Pioro-Ladri{\`{e}}re}, \citenamefont {Brunner}, \citenamefont {Lu},
		\citenamefont {Nakajima}, \citenamefont {Obata}, \citenamefont {Noiri},
		\citenamefont {Palmstr{\o}m}, \citenamefont {Gossard},\ and\ \citenamefont
		{Tarucha}}]{Yoneda2015}%
	\BibitemOpen
	\bibfield  {author} {\bibinfo {author} {\bibfnamefont {J.}~\bibnamefont
			{Yoneda}}, \bibinfo {author} {\bibfnamefont {T.}~\bibnamefont {Otsuka}},
		\bibinfo {author} {\bibfnamefont {T.}~\bibnamefont {Takakura}}, \bibinfo
		{author} {\bibfnamefont {M.}~\bibnamefont {Pioro-Ladri{\`{e}}re}}, \bibinfo
		{author} {\bibfnamefont {R.}~\bibnamefont {Brunner}}, \bibinfo {author}
		{\bibfnamefont {H.}~\bibnamefont {Lu}}, \bibinfo {author} {\bibfnamefont
			{T.}~\bibnamefont {Nakajima}}, \bibinfo {author} {\bibfnamefont
			{T.}~\bibnamefont {Obata}}, \bibinfo {author} {\bibfnamefont
			{A.}~\bibnamefont {Noiri}}, \bibinfo {author} {\bibfnamefont {C.~J.}\
			\bibnamefont {Palmstr{\o}m}}, \bibinfo {author} {\bibfnamefont {A.~C.}\
			\bibnamefont {Gossard}}, \ and\ \bibinfo {author} {\bibfnamefont
			{S.}~\bibnamefont {Tarucha}},\ }\href {\doibase 10.7567/apex.8.084401}
	{\bibfield  {journal} {\bibinfo  {journal} {Applied Physics Express}\
		}\textbf {\bibinfo {volume} {8}},\ \bibinfo {pages} {084401} (\bibinfo {year}
		{2015})}\BibitemShut {NoStop}%
	\bibitem [{\citenamefont {Hanson}\ and\ \citenamefont
		{Burkard}(2007)}]{Hanson2007b}%
	\BibitemOpen
	\bibfield  {author} {\bibinfo {author} {\bibfnamefont {R.}~\bibnamefont
			{Hanson}}\ and\ \bibinfo {author} {\bibfnamefont {G.}~\bibnamefont
			{Burkard}},\ }\href {\doibase 10.1103/PhysRevLett.98.050502} {\bibfield
		{journal} {\bibinfo  {journal} {Phys. Rev. Lett.}\ }\textbf {\bibinfo
			{volume} {98}},\ \bibinfo {pages} {050502} (\bibinfo {year}
		{2007})}\BibitemShut {NoStop}%
	\bibitem [{\citenamefont {Bodenstedt}\ \emph {et~al.}(2018)\citenamefont
		{Bodenstedt}, \citenamefont {Jakobi}, \citenamefont {Michl}, \citenamefont
		{Gerhardt}, \citenamefont {Neumann},\ and\ \citenamefont
		{Wrachtrup}}]{1804.02893}%
	\BibitemOpen
	\bibfield  {author} {\bibinfo {author} {\bibfnamefont {S.}~\bibnamefont
			{Bodenstedt}}, \bibinfo {author} {\bibfnamefont {I.}~\bibnamefont {Jakobi}},
		\bibinfo {author} {\bibfnamefont {J.}~\bibnamefont {Michl}}, \bibinfo
		{author} {\bibfnamefont {I.}~\bibnamefont {Gerhardt}}, \bibinfo {author}
		{\bibfnamefont {P.}~\bibnamefont {Neumann}}, \ and\ \bibinfo {author}
		{\bibfnamefont {J.}~\bibnamefont {Wrachtrup}},\ }\href@noop {} {\enquote
		{\bibinfo {title} {Nanoscale spin manipulation with pulsed magnetic gradient
				fields from a hard disc drive writer},}\ } (\bibinfo {year} {2018}),\ \Eprint
	{http://arxiv.org/abs/arXiv:1804.02893} {arXiv:1804.02893} \BibitemShut
	{NoStop}%
	\bibitem [{\citenamefont {Redfield}(1957)}]{Redfield1957}%
	\BibitemOpen
	\bibfield  {author} {\bibinfo {author} {\bibfnamefont {A.~G.}\ \bibnamefont
			{Redfield}},\ }\href {\doibase 10.1147/rd.11.0019} {\bibfield  {journal}
		{\bibinfo  {journal} {{IBM} Journal of Research and Development}\ }\textbf
		{\bibinfo {volume} {1}},\ \bibinfo {pages} {19} (\bibinfo {year}
		{1957})}\BibitemShut {NoStop}%
	\bibitem [{\citenamefont {Kohler}\ and\ \citenamefont
		{H\"{a}nggi}(2006)}]{Kohler2006}%
	\BibitemOpen
	\bibfield  {author} {\bibinfo {author} {\bibfnamefont {S.}~\bibnamefont
			{Kohler}}\ and\ \bibinfo {author} {\bibfnamefont {P.}~\bibnamefont
			{H\"{a}nggi}},\ }\href {\doibase 10.1002/prop.200610314} {\bibfield
		{journal} {\bibinfo  {journal} {Fortschritte der Physik}\ }\textbf {\bibinfo
			{volume} {54}},\ \bibinfo {pages} {804} (\bibinfo {year} {2006})}\BibitemShut
	{NoStop}%
	\bibitem [{\citenamefont {Qi}\ \emph {et~al.}(2017)\citenamefont {Qi},
		\citenamefont {Wu}, \citenamefont {Ward}, \citenamefont {Prance},
		\citenamefont {Kim}, \citenamefont {Gamble}, \citenamefont {Mohr},
		\citenamefont {Shi}, \citenamefont {Savage}, \citenamefont {Lagally},
		\citenamefont {Eriksson}, \citenamefont {Friesen}, \citenamefont
		{Coppersmith},\ and\ \citenamefont {Vavilov}}]{Qi2017}%
	\BibitemOpen
	\bibfield  {author} {\bibinfo {author} {\bibfnamefont {Z.}~\bibnamefont
			{Qi}}, \bibinfo {author} {\bibfnamefont {X.}~\bibnamefont {Wu}}, \bibinfo
		{author} {\bibfnamefont {D.~R.}\ \bibnamefont {Ward}}, \bibinfo {author}
		{\bibfnamefont {J.~R.}\ \bibnamefont {Prance}}, \bibinfo {author}
		{\bibfnamefont {D.}~\bibnamefont {Kim}}, \bibinfo {author} {\bibfnamefont
			{J.~K.}\ \bibnamefont {Gamble}}, \bibinfo {author} {\bibfnamefont {R.~T.}\
			\bibnamefont {Mohr}}, \bibinfo {author} {\bibfnamefont {Z.}~\bibnamefont
			{Shi}}, \bibinfo {author} {\bibfnamefont {D.~E.}\ \bibnamefont {Savage}},
		\bibinfo {author} {\bibfnamefont {M.~G.}\ \bibnamefont {Lagally}}, \bibinfo
		{author} {\bibfnamefont {M.~A.}\ \bibnamefont {Eriksson}}, \bibinfo {author}
		{\bibfnamefont {M.}~\bibnamefont {Friesen}}, \bibinfo {author} {\bibfnamefont
			{S.~N.}\ \bibnamefont {Coppersmith}}, \ and\ \bibinfo {author} {\bibfnamefont
			{M.~G.}\ \bibnamefont {Vavilov}},\ }\href {\doibase
		10.1103/PhysRevB.96.115305} {\bibfield  {journal} {\bibinfo  {journal} {Phys.
				Rev. B}\ }\textbf {\bibinfo {volume} {96}},\ \bibinfo {pages} {115305}
		(\bibinfo {year} {2017})}\BibitemShut {NoStop}%
	\bibitem [{\citenamefont {Makhlin}\ and\ \citenamefont
		{Shnirman}(2004)}]{Makhlin2004}%
	\BibitemOpen
	\bibfield  {author} {\bibinfo {author} {\bibfnamefont {Y.}~\bibnamefont
			{Makhlin}}\ and\ \bibinfo {author} {\bibfnamefont {A.}~\bibnamefont
			{Shnirman}},\ }\href {\doibase 10.1103/PhysRevLett.92.178301} {\bibfield
		{journal} {\bibinfo  {journal} {Phys. Rev. Lett.}\ }\textbf {\bibinfo
			{volume} {92}},\ \bibinfo {pages} {178301} (\bibinfo {year}
		{2004})}\BibitemShut {NoStop}%
	\bibitem [{\citenamefont {Dial}\ \emph {et~al.}(2013)\citenamefont {Dial},
		\citenamefont {Shulman}, \citenamefont {Harvey}, \citenamefont {Bluhm},
		\citenamefont {Umansky},\ and\ \citenamefont {Yacoby}}]{Dial2013}%
	\BibitemOpen
	\bibfield  {author} {\bibinfo {author} {\bibfnamefont {O.~E.}\ \bibnamefont
			{Dial}}, \bibinfo {author} {\bibfnamefont {M.~D.}\ \bibnamefont {Shulman}},
		\bibinfo {author} {\bibfnamefont {S.~P.}\ \bibnamefont {Harvey}}, \bibinfo
		{author} {\bibfnamefont {H.}~\bibnamefont {Bluhm}}, \bibinfo {author}
		{\bibfnamefont {V.}~\bibnamefont {Umansky}}, \ and\ \bibinfo {author}
		{\bibfnamefont {A.}~\bibnamefont {Yacoby}},\ }\href {\doibase
		10.1103/PhysRevLett.110.146804} {\bibfield  {journal} {\bibinfo  {journal}
			{Phys. Rev. Lett.}\ }\textbf {\bibinfo {volume} {110}},\ \bibinfo {pages}
		{146804} (\bibinfo {year} {2013})}\BibitemShut {NoStop}%
	\bibitem [{\citenamefont {Wu}\ \emph {et~al.}(2014)\citenamefont {Wu},
		\citenamefont {Ward}, \citenamefont {Prance}, \citenamefont {Kim},
		\citenamefont {Gamble}, \citenamefont {Mohr}, \citenamefont {Shi},
		\citenamefont {Savage}, \citenamefont {Lagally}, \citenamefont {Friesen},
		\citenamefont {Coppersmith},\ and\ \citenamefont {Eriksson}}]{Wu2014}%
	\BibitemOpen
	\bibfield  {author} {\bibinfo {author} {\bibfnamefont {X.}~\bibnamefont
			{Wu}}, \bibinfo {author} {\bibfnamefont {D.~R.}\ \bibnamefont {Ward}},
		\bibinfo {author} {\bibfnamefont {J.~R.}\ \bibnamefont {Prance}}, \bibinfo
		{author} {\bibfnamefont {D.}~\bibnamefont {Kim}}, \bibinfo {author}
		{\bibfnamefont {J.~K.}\ \bibnamefont {Gamble}}, \bibinfo {author}
		{\bibfnamefont {R.~T.}\ \bibnamefont {Mohr}}, \bibinfo {author}
		{\bibfnamefont {Z.}~\bibnamefont {Shi}}, \bibinfo {author} {\bibfnamefont
			{D.~E.}\ \bibnamefont {Savage}}, \bibinfo {author} {\bibfnamefont {M.~G.}\
			\bibnamefont {Lagally}}, \bibinfo {author} {\bibfnamefont {M.}~\bibnamefont
			{Friesen}}, \bibinfo {author} {\bibfnamefont {S.~N.}\ \bibnamefont
			{Coppersmith}}, \ and\ \bibinfo {author} {\bibfnamefont {M.~A.}\ \bibnamefont
			{Eriksson}},\ }\href {\doibase 10.1073/pnas.1412230111} {\bibfield  {journal}
		{\bibinfo  {journal} {Proceedings of the National Academy of Sciences}\
		}\textbf {\bibinfo {volume} {111}},\ \bibinfo {pages} {11938} (\bibinfo
		{year} {2014})}\BibitemShut {NoStop}%
	\bibitem [{\citenamefont {Li}\ \emph {et~al.}(2017)\citenamefont {Li},
		\citenamefont {Barnes}, \citenamefont {Kestner},\ and\ \citenamefont
		{Das~Sarma}}]{Li2017}%
	\BibitemOpen
	\bibfield  {author} {\bibinfo {author} {\bibfnamefont {X.}~\bibnamefont
			{Li}}, \bibinfo {author} {\bibfnamefont {E.}~\bibnamefont {Barnes}}, \bibinfo
		{author} {\bibfnamefont {J.~P.}\ \bibnamefont {Kestner}}, \ and\ \bibinfo
		{author} {\bibfnamefont {S.}~\bibnamefont {Das~Sarma}},\ }\href {\doibase
		10.1103/PhysRevA.96.012309} {\bibfield  {journal} {\bibinfo  {journal} {Phys.
				Rev. A}\ }\textbf {\bibinfo {volume} {96}},\ \bibinfo {pages} {012309}
		(\bibinfo {year} {2017})}\BibitemShut {NoStop}%
	\bibitem [{\citenamefont {Russ}\ \emph {et~al.}(2016)\citenamefont {Russ},
		\citenamefont {Ginzel},\ and\ \citenamefont {Burkard}}]{Russ2016}%
	\BibitemOpen
	\bibfield  {author} {\bibinfo {author} {\bibfnamefont {M.}~\bibnamefont
			{Russ}}, \bibinfo {author} {\bibfnamefont {F.}~\bibnamefont {Ginzel}}, \ and\
		\bibinfo {author} {\bibfnamefont {G.}~\bibnamefont {Burkard}},\ }\href
	{\doibase 10.1103/PhysRevB.94.165411} {\bibfield  {journal} {\bibinfo
			{journal} {Phys. Rev. B}\ }\textbf {\bibinfo {volume} {94}},\ \bibinfo
		{pages} {165411} (\bibinfo {year} {2016})}\BibitemShut {NoStop}%
	\bibitem [{\citenamefont {Bello}\ \emph {et~al.}(2017)\citenamefont {Bello},
		\citenamefont {Platero},\ and\ \citenamefont {Kohler}}]{Bello2017}%
	\BibitemOpen
	\bibfield  {author} {\bibinfo {author} {\bibfnamefont {M.}~\bibnamefont
			{Bello}}, \bibinfo {author} {\bibfnamefont {G.}~\bibnamefont {Platero}}, \
		and\ \bibinfo {author} {\bibfnamefont {S.}~\bibnamefont {Kohler}},\ }\href
	{\doibase 10.1103/PhysRevB.96.045408} {\bibfield  {journal} {\bibinfo
			{journal} {Phys. Rev. B}\ }\textbf {\bibinfo {volume} {96}},\ \bibinfo
		{pages} {045408} (\bibinfo {year} {2017})}\BibitemShut {NoStop}%
	\bibitem [{\citenamefont {Danon}\ and\ \citenamefont
		{Nazarov}(2009)}]{Danon2009}%
	\BibitemOpen
	\bibfield  {author} {\bibinfo {author} {\bibfnamefont {J.}~\bibnamefont
			{Danon}}\ and\ \bibinfo {author} {\bibfnamefont {Y.~V.}\ \bibnamefont
			{Nazarov}},\ }\href {\doibase 10.1103/PhysRevB.80.041301} {\bibfield
		{journal} {\bibinfo  {journal} {Phys. Rev. B}\ }\textbf {\bibinfo {volume}
			{80}},\ \bibinfo {pages} {041301} (\bibinfo {year} {2009})}\BibitemShut
	{NoStop}%
	\bibitem [{\citenamefont {Yang}\ \emph {et~al.}(2013)\citenamefont {Yang},
		\citenamefont {Rossi}, \citenamefont {Ruskov}, \citenamefont {Lai},
		\citenamefont {Mohiyaddin}, \citenamefont {Lee}, \citenamefont {Tahan},
		\citenamefont {Klimeck}, \citenamefont {Morello},\ and\ \citenamefont
		{Dzurak}}]{Yang2013}%
	\BibitemOpen
	\bibfield  {author} {\bibinfo {author} {\bibfnamefont {C.~H.}\ \bibnamefont
			{Yang}}, \bibinfo {author} {\bibfnamefont {A.}~\bibnamefont {Rossi}},
		\bibinfo {author} {\bibfnamefont {R.}~\bibnamefont {Ruskov}}, \bibinfo
		{author} {\bibfnamefont {N.~S.}\ \bibnamefont {Lai}}, \bibinfo {author}
		{\bibfnamefont {F.~A.}\ \bibnamefont {Mohiyaddin}}, \bibinfo {author}
		{\bibfnamefont {S.}~\bibnamefont {Lee}}, \bibinfo {author} {\bibfnamefont
			{C.}~\bibnamefont {Tahan}}, \bibinfo {author} {\bibfnamefont
			{G.}~\bibnamefont {Klimeck}}, \bibinfo {author} {\bibfnamefont
			{A.}~\bibnamefont {Morello}}, \ and\ \bibinfo {author} {\bibfnamefont
			{A.~S.}\ \bibnamefont {Dzurak}},\ }\href {\doibase 10.1038/ncomms3069}
	{\bibfield  {journal} {\bibinfo  {journal} {Nature Communications}\ }\textbf
		{\bibinfo {volume} {4}} (\bibinfo {year} {2013}),\
		10.1038/ncomms3069}\BibitemShut {NoStop}%
	\bibitem [{\citenamefont {Veldhorst}\ \emph {et~al.}(2015)\citenamefont
		{Veldhorst}, \citenamefont {Ruskov}, \citenamefont {Yang}, \citenamefont
		{Hwang}, \citenamefont {Hudson}, \citenamefont {Flatt\'e}, \citenamefont
		{Tahan}, \citenamefont {Itoh}, \citenamefont {Morello},\ and\ \citenamefont
		{Dzurak}}]{Veldhorst2015}%
	\BibitemOpen
	\bibfield  {author} {\bibinfo {author} {\bibfnamefont {M.}~\bibnamefont
			{Veldhorst}}, \bibinfo {author} {\bibfnamefont {R.}~\bibnamefont {Ruskov}},
		\bibinfo {author} {\bibfnamefont {C.~H.}\ \bibnamefont {Yang}}, \bibinfo
		{author} {\bibfnamefont {J.~C.~C.}\ \bibnamefont {Hwang}}, \bibinfo {author}
		{\bibfnamefont {F.~E.}\ \bibnamefont {Hudson}}, \bibinfo {author}
		{\bibfnamefont {M.~E.}\ \bibnamefont {Flatt\'e}}, \bibinfo {author}
		{\bibfnamefont {C.}~\bibnamefont {Tahan}}, \bibinfo {author} {\bibfnamefont
			{K.~M.}\ \bibnamefont {Itoh}}, \bibinfo {author} {\bibfnamefont
			{A.}~\bibnamefont {Morello}}, \ and\ \bibinfo {author} {\bibfnamefont
			{A.~S.}\ \bibnamefont {Dzurak}},\ }\href {\doibase
		10.1103/PhysRevB.92.201401} {\bibfield  {journal} {\bibinfo  {journal} {Phys.
				Rev. B}\ }\textbf {\bibinfo {volume} {92}},\ \bibinfo {pages} {201401}
		(\bibinfo {year} {2015})}\BibitemShut {NoStop}%
\end{thebibliography}
\end{document}

\title{Supplementary information for ``Coherent long-range transfer of
two-electron states in ac driven triple quantum dots''}

\author{Jordi Pic\'o-Cort\'es}
\email{jordi.p.cortes@csic.es}
\affiliation{Materials Science Factory, Instituto de Ciencia de Materiales de
	Madrid (CSIC), 28049, Madrid, Spain }
\affiliation{Institute for Theoretical Physics, University of Regensburg, 93040 Regensburg, Germany}

\author{Fernando Gallego-Marcos}

\author{Gloria Platero}
\affiliation{Materials Science Factory, Instituto de Ciencia de Materiales de
	Madrid (CSIC), 28049, Madrid, Spain }
\maketitle

\section{Effective Hamiltonian}

In this section, we give the expressions for the different energies
and rates appearing in the effective Hamiltonian in the main text

\begin{align}
 & \tilde{E}_{\ket{\sigma,\sigma',0}}(t)=E_{\text{LC}}-s\Delta_{\mathrm{LC}}-\sum_{\nu,\mu}\frac{t_{\text{CR}}^{\nu*}(t)t_{\text{CR}}^{\mu}(t)}{\delta_{101}+s\Delta_{\mathrm{LC}}+\mu\hbar\omega}\label{eq:ELC}\\
 & +\sum_{\nu,\mu}\left[\frac{t_{\text{LC}}^{\nu*}(t)t_{\text{LC}}^{\mu}(t)}{\delta_{020}+s\Delta_{\mathrm{LC}}-\nu\hbar\omega}+\frac{t_{\text{LC}}^{\nu*}(t)t_{\text{LC}}^{\mu}(t)}{\delta_{200}+s\Delta_{\mathrm{LC}}-\nu\hbar\omega}\right],\nonumber 
\end{align}
\begin{align}
 & \tilde{E}_{\ket{0,\sigma,\sigma'}}(t)=E_{\text{CR}}-s\Delta_{\mathrm{CR}}-\sum_{\nu,\mu}\frac{t_{\text{LC}}^{\nu*}(t)t_{\text{LC}}^{\mu}(t)}{\zeta_{101}+s\Delta_{\mathrm{CR}}+\mu\hbar\omega}\label{eq:ECR}\\
 & +\sum_{\nu,\mu}\left[\frac{t_{\text{CR}}^{\nu*}(t)t_{\text{CR}}^{\mu*}(t)}{\zeta_{020}+s\Delta_{\mathrm{CR}}-\nu\hbar\omega}+\frac{t_{\text{CR}}^{\nu*}(t)t_{\text{CR}}^{\mu*}(t)}{\zeta_{002}+s\Delta_{\mathrm{CR}}-\nu\hbar\omega}\right],\nonumber 
\end{align}
where $\sigma\neq\sigma'$ and $s=\pm1$ for $\sigma=\uparrow,\downarrow$;
the indices $\nu$ and $\mu$ indicate the sideband numbers. Unless
explicitly noted, $\nu$ and $\mu$ go from $-\infty$ to $\infty$.
The virtual tunneling rates are 
\begin{align}
\tau_{110}(t) & =\sum_{\mu,\nu}\frac{t_{\mathrm{LC}}^{\mu*}(t)t_{\mathrm{LC}}^{\nu}(t)}{2}\left[\frac{1}{\delta_{020}+\Delta_{\mathrm{LC}}-\nu\hbar\omega}+\frac{1}{\delta_{200}+\Delta_{\mathrm{LC}}-\nu\hbar\omega}+\frac{1}{\delta_{020}-\Delta_{\mathrm{LC}}-\nu\hbar\omega}+\frac{1}{\delta_{200}-\Delta_{\mathrm{LC}}-\nu\hbar\omega}\right],\label{eq:tau110}\\
\tau_{011}(t) & =\sum_{\mu,\nu}\frac{t_{\mathrm{CR}}^{\mu*}(t)t_{\mathrm{CR}}^{\nu}(t)}{2}\left[\frac{1}{\zeta_{020}+\Delta_{\mathrm{CR}}-\nu\hbar\omega}+\frac{1}{\zeta_{002}+\Delta_{\mathrm{CR}}-\nu\hbar\omega}+\frac{1}{\zeta_{020}-\Delta_{\mathrm{CR}}-\nu\hbar\omega}+\frac{1}{\zeta_{002}-\Delta_{\mathrm{CR}}-\nu\hbar\omega}\right],\label{eq:tau011}
\end{align}
\begin{align}
\tau_{\text{LR},1}(t) & =\sum_{\nu,\mu}\frac{t_{\text{LC}}^{\nu*}(t)t_{\text{CR}}^{\mu}(t)}{2}\left[\frac{1}{\delta_{020}+\Delta_{\mathrm{LC}}-\nu\hbar\omega}+\frac{1}{\zeta_{020}+\Delta_{\mathrm{CR}}-\mu\hbar\omega}-\frac{1}{\delta_{101}+\Delta_{\mathrm{CR}}+\mu\hbar\omega}-\frac{1}{\zeta_{101}+\Delta_{\mathrm{LC}}+\nu\hbar\omega}\right],\label{eq:tLR1}
\end{align}
\begin{align}
\tau_{\text{LR},2}(t) & =-\sum_{\nu,\mu}\frac{t_{\text{LC}}^{\nu*}(t)t_{\text{CR}}^{\mu}(t)}{2}\left[\frac{1}{\delta_{020}+\Delta_{\mathrm{LC}}-\nu\hbar\omega}+\frac{1}{\zeta_{020}+\Delta_{\mathrm{CR}}-\mu\hbar\omega}\right].\label{eq:tLR2}
\end{align}

In the RWA approximation, the above expressions read

\begin{align}
 & \tilde{E}_{\ket{\sigma,\sigma',0}}^{\mathrm{RWA}}=E_{\text{LC}}-s\Delta_{\mathrm{LC}}-\sum_{\nu}\frac{|t_{\text{CR}}^{\nu}|^{2}}{\delta_{101}+s\Delta_{\mathrm{LC}}+\mu\hbar\omega}\label{eq:ELC-RWA}\\
 & +\sum_{\nu}\left[\frac{|t_{\text{LC}}^{\nu}|^{2}}{\delta_{020}+s\Delta_{\mathrm{LC}}-\nu\hbar\omega}+\frac{|t_{\text{LC}}^{\nu}|^{2}}{\delta_{200}+s\Delta_{\mathrm{LC}}-\nu\hbar\omega}\right],\nonumber 
\end{align}
\begin{align}
 & \tilde{E}_{\ket{0,\sigma,\sigma'}}^{\mathrm{RWA}}=E_{\text{CR}}-s\Delta_{\mathrm{CR}}-\sum_{\nu}\frac{|t_{\text{LC}}^{\nu}|^{2}}{\zeta_{101}+s\Delta_{\mathrm{CR}}+\mu\hbar\omega}\label{eq:ECR-RWA}\\
 & +\sum_{\nu}\left[\frac{|t_{\text{CR}}^{\nu}|^{2}}{\zeta_{020}+s\Delta_{\mathrm{CR}}-\nu\hbar\omega}+\frac{|t_{\text{CR}}^{\nu}|^{2}}{\zeta_{002}+s\Delta_{\mathrm{CR}}-\nu\hbar\omega}\right],\nonumber 
\end{align}

\begin{align}
\tau_{110}^{\mathrm{RWA}} & =\sum_{\nu}\frac{|t_{\text{LC}}^{\nu}|^{2}}{2}\left[\frac{1}{\delta_{020}+\Delta_{\mathrm{LC}}-\nu\hbar\omega}+\frac{1}{\delta_{200}+\Delta_{\mathrm{LC}}-\nu\hbar\omega}+\frac{1}{\delta_{020}-\Delta_{\mathrm{LC}}-\nu\hbar\omega}+\frac{1}{\delta_{200}-\Delta_{\mathrm{LC}}-\nu\hbar\omega}\right],\label{eq:tau110-RWA}\\
\tau_{011}^{\mathrm{RWA}} & =\sum_{\nu}\frac{|t_{\text{CR}}^{\nu}|^{2}}{2}\left[\frac{1}{\zeta_{020}+\Delta_{\mathrm{CR}}-\nu\hbar\omega}+\frac{1}{\zeta_{002}+\Delta_{\mathrm{CR}}-\nu\hbar\omega}+\frac{1}{\zeta_{020}-\Delta_{\mathrm{CR}}-\nu\hbar\omega}+\frac{1}{\zeta_{002}-\Delta_{\mathrm{CR}}-\nu\hbar\omega}\right],\label{eq:tau011-RWA}
\end{align}
\begin{align}
 & \tau_{\text{LR},1}^{\mathrm{RWA},n}=\sum_{\nu}\frac{\tau_{\mathrm{LC}}\tau_{\mathrm{CR}}}{2}J_{\nu}\left(\frac{V_{\mathrm{L}}^{\mathrm{ac}}}{\hbar\omega}\right)J_{\nu-n}\left(\frac{V_{\mathrm{R}}^{\mathrm{ac}}}{\hbar\omega}\right)\nonumber \\
 & \times\left[\frac{1}{\delta_{020}+\Delta_{\mathrm{LC}}-\nu\hbar\omega}+\frac{1}{\zeta_{020}+\Delta_{\mathrm{CR}}-(\nu-n)\hbar\omega}-\frac{1}{\delta_{101}+\Delta_{\mathrm{CR}}+(\nu-n)\hbar\omega}-\frac{1}{\zeta_{101}+\Delta_{\mathrm{LC}}+\nu\hbar\omega},\right]\label{eq:tLR1-RWA}
\end{align}
\begin{align}
\tau_{\text{LR},2}^{\mathrm{RWA},n}= & -\sum_{\nu}\frac{\tau_{\mathrm{LC}}\tau_{\mathrm{CR}}}{2}J_{\nu}\left(\frac{V_{\mathrm{L}}^{\mathrm{ac}}}{\hbar\omega}\right)J_{\nu-n}\left(\frac{V_{\mathrm{R}}^{\mathrm{ac}}}{\hbar\omega}\right)\left[\frac{1}{\delta_{020}+\Delta_{\mathrm{LC}}-\nu\hbar\omega}+\frac{1}{\zeta_{020}+\Delta_{\mathrm{CR}}-(\nu-n)\hbar\omega}\right].\label{eq:tLR2-RWA}
\end{align}

In the main text and the following sections we have neglected to include
the $\mathrm{RWA}$ labels. All formulas are based on the RWA approximation
unless explicitly noted.

\section{Entanglement Measure}

To measure the entanglement of the prepared  state we use the expression
$\mathcal{E}(\mathcal{C})=\mathcal{S}_{2}\left(1/2+1/2\sqrt{1-\mathcal{C}^{2}}\right)$\cite{Hill1997,Wootters1998}
where $S_{2}(x)$ is the binary entropy function: $S_{2}(x)=-[x\text{log}_{2}(x)+(1-x)\text{log}_{2}(1-x)]$
and $\mathcal{C}$ is the concurrence. Both concurrence and entanglement
range from zero to one and the concurrence is monotonically related
to the entanglement, hence it is also a measure of entanglement. The
expression for $\mathcal{C}$ is: $\mathcal{C}=\text{max}\{0,\sqrt{\lambda_{1}}-\sqrt{\lambda_{2}}-\sqrt{\lambda_{3}}-\sqrt{\lambda_{4}}\}$,
where $\lambda_{i}$ are the sorted eigenvalues of the matrix $R_{\mathcal{C}}=\rho(\sigma_{y}\otimes\sigma_{y})\rho(\sigma_{y}\otimes\sigma_{y})$
and $\rho$ is the density matrix of the four different states of
two particles; for the left and center dots: ${\ket{\up,\up,0},\ket{\up,\dw,0},\ket{\dw,\up,0},\ket{\dw,\dw,0}}$. 

\section{General Transfer Protocol}

Under the sideband interference condition (i.e: $\tau_{110}=0$, $\tau_{011}=0$
and $\tau_{\mathrm{LR,2}}=0$), the effective Hamiltonian for the
transfer process is given by 

\begin{equation}
H_{\mathrm{tr}}=\left[\begin{array}{cccc}
\tilde{E}_{|\uparrow,\downarrow,0\mathcal{i}} & 0 & \tau_{\mathrm{LR,1}} & 0\\
0 & \tilde{E}_{|\downarrow,\uparrow,0\mathcal{i}} & 0 & \tau_{\mathrm{LR,1}}\\
\tau_{\mathrm{LR,1}} & 0 & \tilde{E}_{|0,\uparrow,\downarrow\mathcal{i}}-n\hbar\omega & 0\\
0 & \tau_{\mathrm{LR,1}} & 0 & \tilde{E}_{|0,\downarrow,\uparrow\mathcal{i}}-n\hbar\omega
\end{array}\right]\label{eq:Htr}
\end{equation}

The energy of the four states are also renormalized by the virtual
tunneling processes. We will consider $n=0$, in which case the situation
is much simplified by the symmetries between $\mathcal{Q}_{L}$ and
$\mathcal{Q}_{R}$ and we can write $\tilde{E}_{|\uparrow,\downarrow,0\mathcal{i}}\simeq-\Delta_{\mathrm{LC}}$,
$\tilde{E}_{|\downarrow,\uparrow,0\mathcal{i}}\simeq\Delta_{\mathrm{LC}}$,
$\tilde{E}_{|0,\uparrow,\downarrow\mathcal{i}}\simeq-\Delta_{\mathrm{CR}}$
and $\tilde{E}_{|0,\downarrow,\uparrow\mathcal{i}}\simeq\Delta_{\mathrm{CR}}$.
We define the unitary operator, $\mathcal{U}_{\mathrm{tr}}(t)=\exp(-iH_{\mathrm{tr}}t/\hbar)$
that describes the time evolution during the transfer process. We
also define the time-evolution operator when the system is left to
evolve under the gradientes (i.e. with $\tau_{\text{LC}}=\tau_{\text{CR}}=0$),
wich under the conditions of above is given by $\mathcal{U}_{Z}(t)=\exp[-iH_{Z}t/\hbar]$,
where $H_{Z}=-\Delta_{\mathrm{LC}}\hat{\sigma}_{\mathrm{LC}}^{z}-\Delta_{\mathrm{CR}}\hat{\sigma}_{\mathrm{CR}}^{z})$.
The maximum fidelity under a single transfer process,\emph{ $\mathcal{F}_{\mathrm{max}}=\left|\bra{\Psi\left(\mathcal{T}_{\mathrm{LR,1}}^{\Omega}/2\right)}\ket{\Psi_{\mathrm{R}}}\right|$},
where\emph{ }$\ket{\Psi\left(t\right)}=\mathcal{U}_{\mathrm{tr}}\left(t\right)\ket{\Psi_{\text{R}}}$
, can be obtained from this. For instance, taking as initial state
an eigenstate of $H_{Z}$, such as $\ket{\Psi_{\text{R}}}=\ket{\up,\dw,0}$,
\begin{align*}
\ket{\Psi\left(\mathcal{T}_{\mathrm{LR,1}}^{\Omega}/2\right)} & =\frac{\left(\Delta_{\text{LC}}-\Delta_{\text{CR}}\right)}{\sqrt{(\Delta_{\mathrm{CR}}-\Delta_{\mathrm{LC}})^{2}+4\tau_{\mathrm{LR,1}}^{2}}}\ket{\up,\dw,0}-\frac{2\tau_{\text{LR,1}}}{\sqrt{(\Delta_{\mathrm{CR}}-\Delta_{\mathrm{LC}})^{2}+4\tau_{\mathrm{LR,1}}^{2}}}\ket{0,\up,\dw},
\end{align*}

If the initial state is not an eigenstate of $H_{Z}$, then the fidelity
under the transfer process will be reduced by the gradients, requiring
further work to bring the state to $\ket{\Psi_{\text{R}}}$. 

The time evolution operator $\mathcal{U}_{\mathrm{tr}}(t)$ can be
parameterized in terms of three angles $\eta=\arctan[(\Delta_{\mathrm{CR}}-\Delta_{\mathrm{LC}})/2\tau_{\mathrm{LR,1}}]$,
$\chi=\sqrt{(\Delta_{\mathrm{CR}}-\Delta_{\mathrm{LC}})^{2}+4\tau_{\mathrm{LR,1}}^{2}}t/\hbar$,
$\delta=(\Delta_{\mathrm{LC}}+\Delta_{\mathrm{CR}})t/\hbar$. as 

\begin{align}
\mathcal{U}_{\mathrm{tr}}(\eta,\chi,\delta)= & \left[\begin{array}{cccc}
a(-\eta,\chi,\delta) & 0 & b(\eta,\chi,\delta) & 0\\
0 & a(-\eta,\chi,-\delta) & 0 & b(\eta,\chi,-\delta)\\
b(\eta,\chi,\delta) & 0 & a(\eta,\chi,\delta) & 0\\
0 & b(\eta,\chi,-\delta) & 0 & a(\eta,\chi,-\delta)
\end{array}\right]\label{eq:Htr2}
\end{align}
with
\[
a(\eta,\chi,\delta)=e^{\frac{i\delta}{2}}\left[\cos\left(\frac{\chi}{2}\right)+i\sin(\eta)\sin\left(\frac{\chi}{2}\right)\right]
\]
\[
b(\eta,\chi,\delta)=-ie^{\frac{i\delta}{2}}\cos(\eta)\sin\left(\frac{\chi}{2}\right)
\]

For fixed gradients, the angles $\{\eta,\chi,\delta\}$ are not independent
from each other. We also parameterize $\mathcal{U}_{Z}(t)\equiv\mathcal{U}_{Z}(\zeta_{\mathrm{LC}},\zeta_{\mathrm{CR}})$
with $\zeta_{k}=\Delta_{k}t/\hbar$, $k=\mathrm{LC,\,CR}$. In the
same way, $\zeta_{\mathrm{CR}}$ and $\zeta_{\mathrm{LC}}$ are both
determined by the pulse time (i.e. the time in which the system is
left to evolve under the gradients) and therefore are not independent
from each other, either. For instance, for the symmetric configuration
($\Delta_{\mathrm{CR}}=-\Delta_{\mathrm{LC}}$) $\delta=0$ and $\zeta_{\mathrm{CR}}=-\zeta_{\mathrm{LC}}$.
In that case we find the following sequence that takes the state in
$\mathcal{Q}_{L}$ to $\mathcal{Q}_{R}$ as
\[
|\Psi_{\mathrm{R}}\mathcal{i}=\mathcal{U}_{\mathrm{tr}}\left(\frac{\pi}{4},\pi,0\right)\mathcal{U}_{Z}\left(\frac{\pi}{2},-\frac{\pi}{2}\right)\mathcal{U}_{\mathrm{tr}}\left(\frac{\pi}{4},\pi,0\right)|\Psi_{\mathrm{L}}\mathcal{i}.
\]

Its physical meaning is discussed in the main text in detail. Note
that in this configuration the $\mathcal{U}_{Z}(\pi/2,-\pi/2)$ transformation
does not depend on the transfer time, contrary to  $\mathcal{U}_{Z}(\mathcal{T}_{\mathrm{off}})$
in the linear configuration. Another simple case corresponds to $\Delta_{\mathrm{CR}}=0$
and $\Delta_{\mathrm{LC}}\neq0$. In this case, to transfer the state
it is enough to take the sequence
\begin{align*}
|\Psi_{\mathrm{R}}\mathcal{i} & =\mathcal{U}_{\mathrm{tr}}\left(\frac{\pi}{4},\pi,\delta_{tr}\right)\mathcal{U}_{Z}(\pi,0)\mathcal{U}_{\mathrm{tr}}\left(\frac{\pi}{4},\pi,\delta_{\mathrm{tr}}\right)\mathcal{U}_{Z}(\delta_{i},0)|\Psi_{\mathrm{L}}\mathcal{i},
\end{align*}
where $\delta_{i}=-\delta_{\mathrm{tr}}-3\pi/2$ . 

For the general case of $\Delta_{\mathrm{LC}}\neq\Delta_{\mathrm{CR}}$,
we define $\mathcal{U}_{\mathrm{LC}}(\alpha_{\mathrm{LC}},\beta_{\mathrm{LC}},\delta_{\mathrm{LC}})$
and $\mathcal{U}_{\mathrm{CR}}\left(\alpha_{\mathrm{CR}},\beta_{\mathrm{CR}},\delta_{\mathrm{CR}}\right)$
the unitary transformations for the two-level systems in $\mathcal{Q}_{\mathrm{L}}$
and $\mathcal{Q}_{\mathrm{R}}$, respectively (i.e: $\mathcal{U}_{\mathrm{LC}}=\exp(-iH_{\mathrm{LC}}t/\hbar)$)
parameterized as $\alpha_{k}=\arctan\left[\Delta_{k}/\tau_{k}\right]$,
$\beta_{k}=\sqrt{\Delta_{k}^{2}+\tau_{k}^{2}}t$ and $\delta_{k}=\Delta_{k}t$,
with $k=\text{LC},\text{CR}$ and $\tau_{\mathrm{LC}}=\tau_{110}$
and $\tau_{\mathrm{CR}}=\tau_{011}$. The transfer process can be
performed through the sequence

\begin{align}
|\Psi_{\mathrm{R}}\mathcal{i} & =\mathcal{U}_{Z}(\zeta_{\mathrm{LC}}^{f},\zeta_{\mathrm{CR}}^{f})\mathcal{U}_{\mathrm{tr}}\left(\frac{\pi}{4},\pi,\delta_{\mathrm{tr}}\right)\mathcal{U}_{\mathrm{CR}}\left(\alpha_{\mathrm{CR}},\pi,\delta_{\mathrm{CR}}\right)\nonumber \\
 & \times\mathcal{U}_{\mathrm{LC}}\left(\alpha_{\mathrm{LC}},\pi,\delta_{\mathrm{LC}}\right)\mathcal{U}_{Z}(\zeta_{\mathrm{LC}}^{i},\zeta_{\mathrm{CR}}^{i})\mathcal{U}_{\mathrm{tr}}\left(\frac{\pi}{4},\pi,\delta_{\mathrm{tr}}\right)|\Psi_{\mathrm{L}}\mathcal{i},\label{eq:generalTransfer}
\end{align}
where $\zeta_{\mathrm{LC}}^{i}+\delta_{\mathrm{LC}}=\zeta_{\mathrm{CR}}^{i}+\delta_{\mathrm{CR}}+(2N+1)\pi$,
$N\mathcal{2}\mathbb{Z}$ and $\zeta_{\mathrm{CR}}^{f}=-(2\delta_{\mathrm{tr}}+\delta_{\mathrm{LC}}+\delta_{\mathrm{CR}}+\zeta_{\mathrm{LC}}^{i}+\zeta_{\mathrm{CR}}^{i})/2$.
The pulses $\mathcal{U}_{k}\left(\alpha_{k},\pi,\delta_{k}\right)$
are active for a time 
\[
\mathcal{T}_{k}=\frac{\hbar\pi}{\sqrt{\Delta_{k}^{2}+\tau_{k}^{2}}},\,\,\,k=\mathrm{LC,\,CR},
\]
while the $\mathcal{U}_{Z}(\zeta_{\mathrm{LC}}^{f},\zeta_{\mathrm{CR}}^{f})$
pulses are active for a time

\begin{align*}
\mathcal{T}_{Z} & =\frac{\hbar\pi}{\Delta_{\mathrm{LC}}-\Delta_{\mathrm{CR}}}\left[2N+1+\frac{\Delta_{\mathrm{CR}}}{\sqrt{\Delta_{\mathrm{CR}}^{2}+\tau_{011}^{2}}}-\frac{\Delta_{\mathrm{LC}}}{\sqrt{\Delta_{\mathrm{LC}}^{2}+\tau_{110}^{2}}}\right].
\end{align*}

\section{Derivation of the decay rates for the manipulation process}

The time-evolution of the density matrix under the assumptions of
weak coupling and Markovianity is given by the Bloch-Redfield master
equation~
\begin{align}
\dot{\rho}(t) & =-i\hbar^{-1}\left[H_{\mathrm{S}},\rho(t)\right]\nonumber \\
 & -\sum_{i}\left[X_{i},\left[Q_{i},\rho(t)\right]\right]-\sum_{i}\left[X_{i},\left\{ R_{i},\rho(t)\right\} \right]\label{eq:masterEq}
\end{align}
where $\left\{ \,,\,\right\} $ indicates the anti-commutator and
\begin{equation}
Q_{i}=\frac{1}{\pi}\int_{0}^{\infty}d\tau\int_{0}^{\infty}d\omega\mathcal{S}(\omega)\tilde{X}_{i}(\tau,0)\cos(\omega\tau)\label{eq:Qj}
\end{equation}
\begin{equation}
R_{i}=\frac{1}{i\pi}\int_{0}^{\infty}d\tau\int_{0}^{\infty}d\omega\mathcal{J}(\omega)\tilde{X}_{i}(\tau,0)\sin(\omega\tau)\label{eq:Rj}
\end{equation}

We define the propagated system operators as $\tilde{X}_{i}(\tau,\tau')=\mathcal{U}^{\dagger}(\tau,\tau')X_{i}\mathcal{U}(\tau,\tau')$.
Apart from $X_{i}=\hat{c}_{i}^{\dagger}\hat{c}_{i}$, we have to consider
the coupling between system and bath through the virtual tunneling
processes, which depend on the energy differences between the states. 

For the manipulation process, as indicated in the main text, the system
is effectively subjected to a single noise source $\xi_{\mathrm{LC}}=\xi_{\mathrm{C}}-\xi_{\mathrm{L}}$.
Up to first order in the system-bath coupling parameter $\alpha$,
the system Hamiltonian together with the bath coupling term reads

\begin{align}
 & H_{\mathrm{LC}}+H_{\mathrm{SB}}=\left(\Delta\tilde{E}_{\mathrm{LC}}^{(0)}+\xi_{\mathrm{LC}}\Delta\tilde{E}_{\mathrm{LC}}^{(1)}\right)\hat{\sigma}_{\mathrm{LC}}^{z}+\left(\tau_{110}^{(0)}+\xi_{\mathrm{LC}}\tau_{110}^{(1)}\right)\hat{\sigma}_{\mathrm{LC}}^{x}\label{eq:systemBathH}
\end{align}
where $\tau_{110}^{(0)}\equiv\left.\tau_{110}\right|_{\xi_{\mathrm{LC}}=0}$
$\tau_{110}^{(1)}\equiv\left.\partial_{\epsilon_{\mathrm{L}}}\tau_{110}\right|_{\xi_{\mathrm{LC}}=0}$,
and similarly $\Delta\tilde{E}_{\mathrm{LC}}^{(0)}=\left.\Delta\tilde{E}_{\mathrm{LC}}\right|_{\xi_{\mathrm{LC}}=0}$,
$\Delta\tilde{E}_{\mathrm{LC}}^{(1)}=\left.\partial_{\epsilon_{\mathrm{L}}}\Delta\tilde{E}_{\mathrm{LC}}\right|_{\xi_{\mathrm{LC}}=0}$.

In order to evaluate the integrals in Eqs.~\ref{eq:Qj} and~\ref{eq:Rj},
we have to find the $\tilde{X}_{i}(\tau,\tau')$ operators defined
in those equations. With the basis employed to define the effective
Hamiltonian in the main text, the operators have a complicated form.
For that reason, we follow Refs.\cite{Kohler2006,Bello2017} and move
to the eigenbasis of the system Hamiltonian, where the interaction
picture operators $\tilde{X}_{i}(\tau,\tau')$ acquire simple sinusoidal
factors. The bath-coupled operator in the eigenbasis is
\begin{align*}
X= & \Delta\tilde{E}_{\mathrm{LC}}^{(1)}\left[\cos\Theta\hat{\sigma}_{\mathrm{LC}}^{z}+\sin\Theta\hat{\sigma}_{\mathrm{LC}}^{x}\right]+\tau_{110}^{(1)}\left[\cos\Theta\hat{\sigma}_{\mathrm{LC}}^{x}+\sin\Theta\hat{\sigma}_{\mathrm{LC}}^{z}\right]
\end{align*}

Here, $\Theta$ is given by $\tau_{110}^{(0)}\tan\left(\Theta/2\right)=-\Delta_{\mathrm{LC}}+\omega_{0}^{(0)}$
, where $\hbar\omega_{0}^{(0)}=\sqrt{\left(\tau_{110}^{(0)}\right)^{2}+\Delta_{\mathrm{LC}}^{2}}$
is the Rabi frequency of the noiseless system. In the interaction
picture it becomes

\begin{align*}
\tilde{X}(\tau,0)=\left(\Delta\tilde{E}_{\mathrm{LC}}^{(1)}\cos\Theta+\tau_{110}^{(1)}\sin\Theta\right)\hat{\sigma}_{\mathrm{LC}}^{z}\\
+\left(\Delta\tilde{E}_{\mathrm{LC}}^{(1)}\sin\Theta+\tau_{110}^{(1)}\cos\Theta\right)\left[\hat{\sigma}_{\mathrm{LC}}^{x}\cos\left(2\omega_{0}^{(0)}\tau\right)+\hat{\sigma}_{\mathrm{LC}}^{y}\sin\left(2\omega_{0}^{(0)}\tau\right)\right]
\end{align*}

Performing the integrals in Eqs.~\ref{eq:Qj}~and~\ref{eq:Rj}
yields

\begin{align}
Q & =\frac{1}{2}\left(\Delta\tilde{E}_{\mathrm{LC}}^{(1)}\cos\Theta+\tau_{110}^{(1)}\sin\Theta\right)\mathcal{S}(0)\hat{\sigma}_{\mathrm{LC}}^{z}+\frac{1}{2}\left(\Delta\tilde{E}_{\mathrm{LC}}^{(1)}\sin\Theta+\tau_{110}^{(1)}\cos\Theta\right)\mathcal{S}(2\omega_{0}^{(0)})\hat{\sigma}_{\mathrm{LC}}^{x}\label{eq:Qman}\\
R= & \frac{1}{2i}\left(\Delta\tilde{E}_{\mathrm{LC}}^{(1)}\cos\Theta+\tau_{110}^{(1)}\sin\Theta\right)\mathcal{J}(0)\hat{\sigma}_{\mathrm{LC}}^{z}+\frac{1}{2i}\left(\Delta\tilde{E}_{\mathrm{LC}}^{(1)}\sin\Theta+\tau_{110}^{(1)}\cos\Theta\right)\mathcal{J}(2\omega_{0}^{(0)})\hat{\sigma}_{\mathrm{LC}}^{x}\label{eq:Rman}
\end{align}

The decay rates can be obtained from Eq.~\ref{eq:masterEq} with
$Q$ as in Eq.~\ref{eq:Qman}, yielding the following rate for the
decay between the two eigenstates of $H_{\text{LC}}$
\begin{equation}
\Gamma_{\mathrm{LC}}=\mathcal{S}(2\omega_{0}^{(0)})\left(\Delta\tilde{E}_{\mathrm{LC}}^{(1)}\sin\Theta+\tau_{110}^{(1)}\cos\Theta\right)^{2}.\label{eq:decayRatesMan}
\end{equation}

\section{Derivation of the decay rates for the transfer process}

During the transfer process, the system couples to charge noise through
several process: (1) direct coupling to charge noise through the gate
energies $\epsilon_{\mathrm{L}},\epsilon_{\mathrm{C}},\epsilon_{\mathrm{R}}$
(2) through the long-range amplitude $\tau_{\mathrm{LR,1}}$ and (3)
because charge noise disrupts the dark state condition and result
in non-zero values for $\tau_{110}$ and $\tau_{\mathrm{LR,2}}$,
as discussed in the main text. In this section, we obtain the decay
rates for the transfer process. To do so, we follow the steps in the
previous section and calculate the $X$ operators. The transfer Hamiltonian
$H_{\mathrm{tr}}$ defined in Eq.~\ref{eq:Htr} has eigenstates

\begin{align}
|\phi_{\sigma,\sigma'}\mathcal{i} & =\sin\left[\sigma\frac{\Upsilon}{2}+(\sigma'-\sigma)\frac{\pi}{4}\right]|-\sigma,\sigma,0\mathcal{i}+\cos\left[\sigma\frac{\Upsilon}{2}+(\sigma'-\sigma)\frac{\pi}{4}\right]|0,-\sigma,\sigma\mathcal{i}\label{eq:eigenstates-Htr}
\end{align}
where $\sigma,\sigma'=\pm1$ (inside the kets they indicate $\uparrow/\downarrow$)
and $\Upsilon$ is given by

\begin{align}
\tau_{\mathrm{LR,1}}\tan\left(\frac{\Upsilon}{2}\right)=\frac{1}{2}\left(\Delta_{\mathrm{LC}}-\Delta_{\mathrm{CR}}+\Lambda\right).\label{eq:upsilon}
\end{align}

The four eigenstates branch out from the four qubit states $\{\ket{\up,\dw,0},\ket{\dw,\up,0},\ket{0,\up,\dw},\ket{0,\dw,\up}\}$
and become hybridized as $\tau_{\mathrm{LR,1}}$ is increased. Their
eigenvalues are 
\begin{align}
\nu_{\sigma,\sigma'} & =\frac{1}{2}\left(\sigma\lambda+\sigma'\Lambda\right),\nonumber \\
\lambda & =(\Delta_{\mathrm{LC}}+\Delta_{\mathrm{CR}}),\\
\Lambda & =\sqrt{(\Delta_{\mathrm{LC}}-\Delta_{\mathrm{CR}})^{2}+4\tau_{\mathrm{LR,1}}^{2}}\label{eq:eigenvals-Htr}
\end{align}

The system is effectively coupled to three noise sources, $\xi_{\mathrm{LC}},\xi_{\mathrm{CR}},\xi_{\mathrm{LR}}$.
In the system eigenbasis $\{|\phi_{++}\mathcal{i},|\phi_{+-}\mathcal{i},|\phi_{-+}\mathcal{i},|\phi_{--}\mathcal{i}\}$,
the direct coupling to the occupation operators yields

\begin{align}
X_{\mathrm{LR}}^{\prime\mathrm{dir}}=\frac{1}{2}\left(\hat{\tau}_{\sigma'}^{x}\sin\Upsilon+\hat{\tau}_{\sigma}^{z}\otimes\hat{\tau}_{\sigma'}^{z}\cos\Upsilon\right)\label{eq:XprimDir}
\end{align}
where $\hat{\tau}_{\sigma}^{i},\hat{\tau}_{\sigma'}^{i}$ are the
$i$th Pauli matrices in the subspaces defined by the degrees of freedom
$\sigma$ and $\sigma'$, respectively and $\Upsilon$ is defined
in Eq.~\ref{eq:upsilon}. Here and onward we write explicitly in
the subscript of the $X$ operators the noise source to which the
system operator is coupled. The  operator for the long-range $\tau_{\mathrm{LR,1}}$
process has matrix elements

\begin{equation}
X_{\mathrm{LC}}^{\prime\mathrm{LR},1}=\frac{1}{2}\sum_{\nu}t_{\mathrm{LC}}^{\nu}t_{\mathrm{CR}}^{-\nu}\left[\frac{1}{\left(\delta_{020}+\Delta_{\mathrm{LC}}-\nu\hbar\omega\right)^{2}}+\frac{1}{\left(\zeta_{101}+\Delta_{\mathrm{LC}}-\nu\hbar\omega\right)^{2}}\right]\left(\hat{\tau}_{\sigma}^{z}\otimes\hat{\tau}_{\sigma'}^{x}\cos\Upsilon-\hat{\tau}_{\sigma'}^{z}\sin\Upsilon\right)\label{eq:XprimLR1LC}
\end{equation}

\begin{align}
X_{\mathrm{CR}}^{\prime\mathrm{LR},1}=-\frac{1}{2}\sum_{\nu}t_{\mathrm{LC}}^{\nu}t_{\mathrm{CR}}^{-\nu}\left[\frac{1}{\left(\zeta_{020}+\Delta_{\mathrm{CR}}+\nu\hbar\omega\right)^{2}}+\frac{1}{\left(\delta_{101}+\Delta_{\mathrm{CR}}+\nu\hbar\omega\right)^{2}}\right]\left(\hat{\tau}_{\sigma}^{z}\otimes\hat{\tau}_{\sigma'}^{x}\cos\Upsilon-\hat{\tau}_{\sigma'}^{z}\sin\Upsilon\right)\label{eq:XprimLR1CR}
\end{align}

Finally, the operators for the deviations from the sideband interference
conditions are given by

\begin{align}
X_{\mathrm{LC}}^{\prime110} & =\frac{1}{4}\sum_{\nu,s=\pm1}|t_{\mathrm{LC}}^{\nu}|^{2}\left[\frac{1}{\left(\delta_{020}+s\Delta_{\mathrm{LC}}-\nu\hbar\omega\right)^{2}}-\frac{1}{\left(\delta_{200}+s'\Delta_{\mathrm{LC}}+\nu\hbar\omega\right)^{2}}\right]\left(\hat{\tau}_{\sigma}^{y}\otimes\hat{\tau}_{\sigma'}^{y}\cos\Upsilon-\hat{\tau}_{\sigma}^{x}\sin\Upsilon+\hat{\tau}_{\sigma}^{x}\otimes\hat{\tau}_{\sigma'}^{x}\right)\label{eq:Xprim110}
\end{align}

\begin{align}
X_{\mathrm{CR}}^{\prime011} & =-\frac{1}{4}\sum_{\nu,s=\pm1}|t_{\mathrm{CR}}^{\nu}|^{2}\left[\frac{1}{\left(\zeta_{020}+s\Delta_{\mathrm{CR}}-\nu\hbar\omega\right)^{2}}-\frac{1}{\left(\zeta_{002}+s'\Delta_{\mathrm{CR}}+\nu\hbar\omega\right)^{2}}\right]\left(\hat{\tau}_{\sigma}^{y}\otimes\hat{\tau}_{\sigma'}^{y}\cos\Upsilon-\hat{\tau}_{\sigma}^{x}\sin\Upsilon-\hat{\tau}_{\sigma}^{x}\otimes\hat{\tau}_{\sigma'}^{x}\right)\label{eq:Xprim011}
\end{align}

\begin{align}
X_{\mathrm{LC}}^{\prime\mathrm{LR,2}}=-\frac{1}{2}\sum_{\nu}\frac{t_{\mathrm{CR}}^{-\nu}t_{\mathrm{LC}}^{\nu}x_{\sigma}^{x}\otimes\hat{\tau}_{\sigma'}^{z}}{\left(\delta_{020}+\Delta_{\mathrm{LC}}-\nu\hbar\omega\right)^{2}}, & \quad X_{\mathrm{CR}}^{\prime\mathrm{LR,2}}=\frac{1}{2}\sum_{\nu}\frac{t_{\mathrm{CR}}^{-\nu}t_{\mathrm{LC}}^{\nu}\hat{\tau}_{\sigma}^{x}\otimes\hat{\tau}_{\sigma'}^{z}}{\left(\zeta_{020}+\Delta_{\mathrm{CR}}-\nu\hbar\omega\right)^{2}}\label{eq:XprimLR2}
\end{align}

Once transformed to the interaction picture, the $\hat{\tau}_{\sigma,\sigma'}^{x,y}$
operators are rotated as $\hat{\tau}_{\sigma'}^{x,y}\to\left[\hat{\tau}_{\sigma'}^{x,y}\cos\left(\Lambda\tau\right)\pm\hat{\tau}_{\sigma'}^{y,x}\sin\left(\Lambda\tau\right)\right]$
and $\hat{\tau}_{\sigma}^{x,y}\to\left[\hat{\tau}_{\sigma}^{x,y}\cos(\lambda\tau)\pm\hat{\tau}_{\sigma}^{y,x}\sin(\lambda\tau)\right]$
(the $\pm$ sign corresponding to $x,y$ respectively). As discussed
in the main text, the contribution due to the direct coupling is particularly
important. Here, we follow in particular how to obtain the related
Bloch-Radfield operators. Starting from Eq.~\ref{eq:XprimDir}

\[
\tilde{X}_{\mathrm{LR}}^{\prime\mathrm{dir}}\left(\tau,0\right)=\frac{1}{2}\left\{ \left[\hat{\tau}_{\sigma'}^{x}\cos\left(\Lambda\tau\right)+\hat{\tau}_{\sigma'}^{y}\sin\left(\Lambda\tau\right)\right]\sin\Upsilon+\hat{\tau}_{\sigma}^{z}\otimes\hat{\tau}_{\sigma'}^{z}\cos\Upsilon\right\} 
\]

The $Q_{i}$ and $R_{i}$ integral can then be performed as above
to give
\begin{align}
Q_{\text{LR}}^{\text{dir}} & =\frac{1}{4}\left[\mathcal{S}\left(\Lambda\right)\sin\Upsilon\hat{\tau}_{\sigma'}^{x}+\mathcal{S}\left(0\right)\cos\Upsilon\hat{\tau}_{\sigma}^{z}\otimes\hat{\tau}_{\sigma'}^{z}\right]\label{eq:QdirLR}\\
R_{\text{LR}}^{\text{dir}} & =\frac{1}{4i}\left[\mathcal{J}\left(\Lambda\right)\sin\Upsilon\hat{\tau}_{\sigma'}^{x}+\mathcal{J}\left(0\right)\cos\Upsilon\hat{\tau}_{\sigma}^{z}\otimes\hat{\tau}_{\sigma'}^{z}\right]\label{eq:RdirLR}
\end{align}

Due to the $1/f$ noise spectrum, the most important contribution
comes from the terms proportional to $\mathcal{S}\left(0\right)$. 

Finally, we write the relaxation rates coming from the different processes
\begin{align}
\Gamma_{\mathrm{LC}}^{\sigma\sigma';\tau\sigma'}=\mathcal{S}(\lambda)\left(\frac{1}{2}\tau_{\mathrm{110}}^{(1)}\sin\Upsilon\right)^{2},\,\,\,\,\, & \Gamma_{\mathrm{CR}}^{\sigma\sigma';\tau\sigma'}=\mathcal{S}(\lambda)\left(\frac{1}{2}\tau_{\mathrm{011}}^{(1)}\sin\Upsilon\right)^{2},\nonumber \\
\Gamma_{\mathrm{LR}}^{\sigma\sigma';\sigma\tau'}=\mathcal{S}(\Lambda)\left(\frac{1}{2}\sin\Upsilon\right)^{2},\,\,\,\,\, & \Gamma_{\mathrm{LC,CR}}^{\sigma\sigma';\sigma\tau'}=\mathcal{S}(\Lambda)\left(\tau_{\mathrm{LR,1}}^{(1)}\cos\Upsilon\right)^{2},\nonumber \\
\Gamma_{\mathrm{LC}}^{\sigma\sigma';\tau\tau'}=2\mathcal{S}(\Lambda+\sigma\tau\lambda)\left[\frac{1}{2}\tau_{110}^{(1)}\left(\cos\Upsilon+1\right)\right]^{2},\,\,\,\,\, & \Gamma_{\mathrm{CR}}^{\sigma\sigma';\tau\tau'}=2\mathcal{S}(\Lambda+\sigma\tau\lambda)\left[\frac{1}{2}\tau_{\mathrm{011}}^{(1)}\left(\cos\Upsilon-1\right)\right]^{2}.\label{eq:relaxation-rates}
\end{align}

$\Gamma_{k}^{\sigma\sigma';\tau\sigma'}$ connects eigenstates with
the same $\sigma'$; $\Gamma_{k}^{\sigma\sigma';\sigma\tau'}$ connects
eigenstates with the same $\sigma$; and $\Gamma_{k}^{\sigma\tau;\sigma'\tau'}$
connects eigenstates differing in both $\sigma$ and $\sigma'$, respectively.
Here we have abused the notation by writing $x^{(1)}$, since it is
not indicated with respect to which variable $x$ is being derived.
However, since it is related to the particular $\Gamma_{i}$ under
which it appears, this does not lead to any ambiguity.